\documentclass[manuscript,nonacm]{acmart}

\renewcommand\footnotetextcopyrightpermission[1]{}
\AtBeginDocument{%
  }

\usepackage{multicol}
\usepackage{multirow}
\usepackage{tcolorbox}
\usepackage{xcolor}
\usepackage{colortbl}

\usepackage[utf8]{inputenc}
\usepackage{geometry}
\usepackage{xcolor}
\usepackage{listings}
\usepackage{tcolorbox}
\usepackage{tikz}
\usepackage{makecell}

\begin{document}

\title{Understanding the Fundamental Design Decisions of Retrieval-Augmented Generation Systems}

\author{Shengming Zhao}
\email{smzhao25@m.fudan.edu.cn}
\orcid{0009-0007-5035-2206}
\affiliation{%
  \institution{Fudan University}
  \city{Shanghai}
  \state{Shanghai}
  \country{China}
}

\author{Yuchen Shao}
\email{ycshao@stu.ecnu.edu.cn}
\orcid{0009-0009-0414-7521}
\affiliation{%
  \institution{East China Normal University and Shanghai Innovation Institute}
  \city{Shanghai}
  \country{China}
}

\author{Yuheng Huang}
\email{yuhenghuang42@g.ecc.u-tokyo.ac.jp}
\orcid{0000-0003-3666-4020}
\affiliation{%
  \institution{The University of Tokyo}
  \city{Tokyo}
  \country{Japan}
}

\author{Jiayang Song}
\email{jiayang.song@ieee.org}
\orcid{0009-0008-7093-9781}
\affiliation{%
  \institution{Macau University of Science and Technology}
  \city{Macau}
  \country{China}
}

\author{Zhijie Wang}
\email{zhijie.wang@concordia.ca}
\orcid{0000-0003-4559-5426}
\affiliation{%
  \institution{Concordia University}
  \city{Montreal}
  \state{QC}
  \country{Canada}
}

\author{Chengcheng Wan}
\authornote{\textbf{Chengcheng Wan and Lei Ma are the corresponding authors.}}
\email{ccwan@sei.ecnu.edu.cn}
\orcid{0000-0001-9162-9688}
\affiliation{%
  \institution{East China Normal University and Shanghai Innovation Institute}
  \city{Shanghai}
  \country{China}
}

\author{Lei Ma}
\authornotemark[1]
\email{ma.lei@acm.org}
\orcid{0000-0002-8621-2420}
\affiliation{%
  \institution{The University of Tokyo, Japan, and University of Alberta}
  \country{Canada}
  }

\renewcommand{\shortauthors}{Zhao et al.}

\begin{abstract}

Retrieval-Augmented Generation (RAG) has emerged as a critical technique for enhancing large language model (LLM) capabilities. However, practitioners face significant challenges when making RAG deployment decisions. While existing research prioritizes algorithmic innovations, a systematic gap persists in understanding fundamental engineering trade-offs that determine RAG success. We present the first comprehensive study of three universal RAG deployment decisions: whether to deploy RAG, how much information to retrieve, and how to integrate retrieved knowledge effectively.

Through systematic experiments across three LLMs and six datasets spanning question answering and code generation tasks, we reveal critical insights: (1) RAG deployment must be highly selective, with variable recall thresholds and failure modes affecting up to 12.6\% of samples even with perfect documents. (2) Optimal retrieval volume exhibits task-dependent behavior—QA tasks show universal patterns (5-10 documents optimal) while code generation requires scenario-specific optimization. (3) Knowledge integration effectiveness depends on task and model characteristics, with code generation benefiting significantly from prompting methods while question answering shows minimal improvement.

These findings demonstrate that universal RAG strategies prove inadequate. Effective RAG systems require context-aware design decisions based on task characteristics and model capabilities. Our analysis provides evidence-based guidance for practitioners and establishes foundational insights for principled RAG deployment. Our code, data and artifacts are publicly available at \url{https://github.com/ShengmingZ/RAG_Benchmark_Code_QA}.

\end{abstract}




\maketitle

\section{Introduction } %

\label{sec: introduction}

\subsection{Motivation}

Retrieval-Augmented Generation (RAG) has emerged as a transformative technique for Large Language Models (LLMs), addressing their limitations in dynamic knowledge access and response reliability~\cite{fan2024survey, RAG_survey}. 
By integrating parametric memory (the pre-trained LLM's internal representations) with non-parametric memory (external knowledge repositories retrieved in real-time), RAG dynamically grounds generation in relevant contextual information~\cite{RAG}. 
It addresses the knowledge limitation of LLMs, improving their response accuracy, contextual relevance, and factual grounding while reducing hallucinations~\cite{ayala-bechard-2024-reducing, zhao2024retrieval, li2024enhancing}.
RAG demonstrates its effectiveness in a wide range of problem domains, including open-domain question answering~\cite{NQ, joshi2017triviaqa, yang2018hotpotqa}, multi-turn dialogue~\cite{wang2024unims}, semantic code completion~\cite{yang2025empirical, liu2024graphcoder}, medical decision making~\cite{xiong2024benchmarking}, software testing~\cite{wang2025code} and many others.
Due to its model-agnostic architecture, performance gains, and modular deployment characteristics, RAG has rapidly transitioned from research prototypes to become a cornerstone technology for production-level AI systems in both academia and industry~\cite{zhao2024retrieval}.

However, effectively deploying RAG in real-world environments presents non-trivial engineering challenges that extend beyond algorithmic design~\cite{hasan2025engineering, SevenFP, ru2024ragchecker}. While recent work focuses on advancing retrieval mechanisms~\cite{searchBestPracticeRAG, RAG_survey} (\emph{e.g.}, dense passage retrieval and hybrid search) and sophisticated reasoning frameworks~\cite{ir-cot, flare} (\emph{e.g.}, iterative retrieval and recursive decomposition), \emph{all} RAG systems, regardless of their architectural complexity, must confront three universal design decisions that fundamentally determine deployment success:

\textbf{Decision-1: Should RAG be deployed?}
The first and most important decision is judging whether the cost introduced by RAG is worth its benefits. Beyond the substantial computation overhead of RAG, it also introduces significant engineering efforts~\cite{shao2025llms, jiang2025rago}, including infrastructure dependency~\cite{SevenFP}, retriever-generator alignment~\cite{yu2024evaluation}, latency bottlenecks, and ongoing knowledge base maintenance~\cite{RAG_survey}.
Neglecting these complexities carries severe practical risks: industry reports estimate failure rates of 72-80\% in enterprise deployments~\cite{KillsOfRAG, RAGWillFail}, where systems lacking rigorous suitability evaluation may confidently cite improper documents, leading to catastrophic operational failures~\cite{KillsOfRAG}.

\textbf{Decision-2: How much information should be retrieved?} 
Another decision is the information volume parameter ($k$), which controls the number of retrieved documents. This seemingly simple parameter governs the trade-off between completeness, noise, and computational efficiency of RAG algorithms~\cite{hasan2025engineering}.
Misconfiguring $k$ triggers cascading failures—retrieving too few documents often excludes the correct answer, while excessive retrieval introduces distracting noise that degrades accuracy and overwhelms system capacity, directly eroding user trust~\cite{SevenFP, PowerofNoise, AWSContext}.

\textbf{Decision-3: How should retrieved knowledge be integrated?} 
Prompting strategies constitute the interface between information retrieval and response generation. Research demonstrates that minimal prompting modifications can yield substantial performance improvements in LLMs across various tasks~\cite{CoT, few-shot}. 
Within RAG architectures, prompting assumes a multifaceted role: it must simultaneously provide task-specific instructions and govern the integration of retrieved contextual information (\emph{e.g.}, instructional framing, contextual utilization protocols, and relevance signaling).
The stakes of this interface are high: poor prompt design has been shown to cause critical misinterpretations of domain terms~\cite{KillsOfRAG}, whereas systematic optimization can drive significant business value (e.g., 23\% revenue increase~\cite{MakePromptGreatAgain}), underscoring the necessity for principled guidance over trial-and-error~\cite{KillsOfRAG}.
This dual role of prompting fundamentally determines whether external knowledge augments or degrades the generation process~\cite{cuconasu2024tale, ir-cot, wang-etal-2024-rear, li2024acecoder}.

Unfortunately, there is a systematic gap in guiding developers through these fundamental design decisions. Existing research predominantly prioritizes algorithmic innovations over foundational engineering choices, creating a significant theory-practice divide. 
Most studies focus on designing and evaluating advanced retrieval techniques~\cite{li2024acecoder, ir-cot, flare, RAG_survey, yang2025empirical, searchBestPracticeRAG}, addressing algorithmic optimizations rather than the fundamental engineering trade-offs that determine RAG deployment success.
While recent work has begun examining engineering challenges in RAG systems~\cite{shao2025llms, jiang2025rago, hasan2025engineering}, these studies overlook the most fundamental design decisions that practitioners require when confronting whether to deploy RAG, how to configure retrieval volume, and how to integrate retrieved knowledge effectively.

\subsection{Contribution}
To tackle these design decisions, we conduct a systematic empirical study of decision-making strategies for RAG systems.
We conduct systematic experiments across three state-of-the-art LLMs (including both open-source and closed-source ones) and six datasets spanning the natural language tasks of question answering (QA) and the software engineering tasks of code generation.

\medskip
\textbf{RQ1: Should RAG be deployed?}
We systematically evaluate RAG's value proposition under varying retrieval recall conditions across diverse datasets and LLMs. This controlled experiment isolated recall as the primary variable of interest while maintaining optimal document quality in other aspects. The results indicate that RAG deployment should be highly selective due to the significant contextual variation: the retrieval recall thresholds required for RAG to outperform base LLMs range from 0.2 to 1.0 across different contexts.

Performance gains exhibit substantial task dependence. In QA tasks, RAG achieves significant improvements (up to 0.6 accuracy). The tasks involving unfamiliar knowledge (NQ~\cite{NQ}, HotpotQA~\cite{yang2018hotpotqa}) show benefits at low recall thresholds, while those with common knowledge (TriviaQA~\cite{joshi2017triviaqa}) require near-perfect retrieval. Code generation demonstrates limited gains (0.1 to 0.25 pass@1), suggesting marginal value for widely-used APIs but greater potential for third-party libraries and private packages.

Most critically, we discover that RAG systems fail on cases solvable by base LLMs, affecting 12.6\% of samples even with perfect documents due to misinterpretation and improper knowledge utilization. These findings establish that RAG deployment requires careful assessment of task characteristics, model capabilities, and tolerance for failure modes rather than universal adoption.

\medskip
\textbf{RQ2: How much information should be retrieved?}
We employ a comprehensive methodology to understand how retrieval volume affects RAG performance across task types. We first systematically evaluate performance across different document numbers on diverse QA and code generation datasets with different LLMs, using statistical significance testing to identify optimal ranges. We then analyze the correlation between generation perplexity and optimal $k$-values to develop deployment-friendly optimization strategies.

Our analysis reveals a fundamental dichotomy between tasks. For QA tasks, we discover a universal ``sweet spot'' of 5-10 documents that delivers optimal performance across datasets and models. Performance plateaus when there are more than 10 documents. Statistical analysis confirms significant improvements up to $k=10$, with diminishing returns thereafter. Remarkably, generation perplexity serves as a reliable proxy for optimal $k$-selection in QA scenarios, enabling test-free optimization.

In contrast, code generation exhibits highly variable, unpredictable performance. The optimal document number varies dramatically from 1 to more than 16,  depending on specific model-dataset combinations. This instability reflects the inherent complexity of multi-document code synthesis, where small changes in retrieved documents can trigger cascading effects on solution quality. Unlike QA tasks, perplexity proves unreliable for code generation optimization, highlighting the need for task-aware RAG system design strategies for code generation.

\medskip
\textbf{RQ3: How should retrieved knowledge be integrated?}
We investigate four prompting strategies (\textit{Prompt Tuning}, \textit{Thought Generation}, \textit{Decomposition}, \textit{Content Verification}) to understand how retrieved knowledge should be integrated into RAG systems across different task types.
We evaluate two representative methods from each category under standardized conditions across zero-shot and few-shot settings.
 
The results show that the prompting effectiveness depends highly on the alignment between model capability, task characteristics, and method selection. 
Code generation tasks benefit significantly from prompting methods, while QA tasks show minimal benefits regardless of approach.
Model capability determines optimal strategies: weaker models like Llama2-13B require few-shot prompting for code generation (up to 85\% improvement on CoNaLA~\cite{conala}) but experience performance degradation on QA tasks, while advanced models demonstrate flexibility across different prompting strategies.
Dataset complexity creates natural boundaries for improvement potential, with tasks showing varying responsiveness to prompting interventions.
For code generation, content verification methods consistently under-perform while prompt tuning and thought generation methods show varying effectiveness depending on task type.

Most remarkably, we reveal that well-designed prompting can enable weaker models to outperform advanced models in code generation scenarios. We discover that prompting methods create orthogonal problem-solving pathways rather than simply enhancing existing approaches. For example, Chain-of-Thought prompting with Llama2-13B on CoNaLA solves an additional 26.2\% of problems while failing on a substantial 8.3\% portion of previously correct samples, demonstrating that prompting fundamentally changes which problems get solved rather than universally improving performance.
These findings highlight the transformative potential of prompting methods for RAG systems, and challenge the assumption of universal prompting strategies, suggesting that we should optimize based on task and model characteristics when constructing prompts for RAG systems.

\medskip

In summary, this work provides a systematic analysis of fundamental RAG deployment decisions, offering evidence-based guidance to bridge the theory-practice gap. Our findings reveal that effective RAG deployment requires a holistic and context-aware approach: the decision to deploy RAG (RQ1) depends on task characteristics and retrieval quality thresholds; optimal retrieval volume (RQ2) follows universal patterns for QA but requires individual optimization for code generation; and integration strategies (RQ3) must align with model capabilities and task types. This multi-dimensional decision framework challenges one-size-fits-all approaches to RAG system design.

This work will contribute to more principled RAG system engineering practices, providing guidance on when to adopt RAG, optimal retrieval strategies, and effective document-LLM integration approaches. It serves as a starting point for tackling the critical systematic RAG system engineering problem. To facilitate reproducibility, we make our code and data publicly available.\footnote{\url{https://github.com/ShengmingZ/RAG_Benchmark_Code_QA}}

\section{Background }
\label{sec:background}

In this section, we provide an overview of the \textit{retrieval} and \textit{generation} phases in RAG systems, detailing their design and implementation.

\begin{figure}
    \centering
    \includegraphics[trim={4cm, 6.5cm, 8.8cm, 4.5cm}, clip, width=0.6\linewidth]{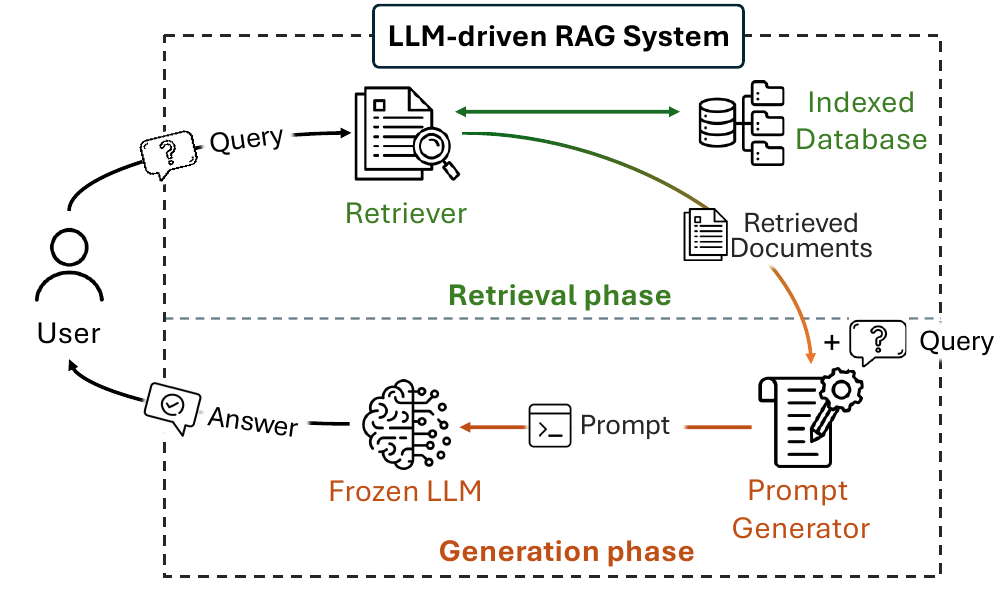}
    \caption{The typical workflow of an LLM-driven RAG system. 
    }
    \label{fig:RAG_workflow}
\end{figure}

\subsection{Retrieval Phase}
As illustrated in Figure~\ref{fig:RAG_workflow}, 
the retrieval phase of an LLM-driven RAG system typically contains two main components: the \textit{indexed database} and the \textit{retriever}.

The \textit{indexed database} is a structured collection of documents $D_1, D_2, \ldots, D_n$, containing domain knowledge and information related to potential user queries. Organized and optimized for efficient searching, it serves as the foundation for retrieving pertinent information.

The \textit{retriever} ranks all documents in the \textit{indexed database} based on their similarity to a query. Given a query $Q$ and the \textit{indexed database} comprising $n$ documents $D_1, D_2, \ldots, D_n$, the \textit{retriever} identifies the order of these documents according to a similarity criterion $sim$. Formally,

\begin{equation}
\footnotesize
\{ D_{i_1}, D_{i_2}, \dots, D_{i_n} \} = \text{sort} \left( \{ D_1, D_2, \dots, D_n \}, \text{sim}(Q, D_i) \right)
\end{equation}

For example, consider a user query $Q$: \textit{``Which is a flowering plant, Pueraria or Pleiospilos?''}. The retriever searches the database to find the most relevant documents ($D_{i_1} \dots D_{i_k}$). In this case, it might retrieve a high-ranking document defining \textit{Pleiospilos} (``...genus of succulent flowering plants...'') alongside distractor documents describing \textit{Pueraria} (``...genus of 15–20 species of plants...'').

There are two primary types of retrievers, based on their use of sparse or dense representations of queries and documents~\cite{SparseDense}.
Sparse retrievers compute similarity by projecting the query and documents into a sparse vector space that aligns with the vocabulary of the documents, typically using traditional Bag-of-Words methods such as TF-IDF or BM25~\cite{BM25}.
To overcome the limitations of these methods, which may struggle with synonyms and varying contextual meanings, dense retrievers~\cite{DPR, contriever} compute similarity scores by encoding queries and documents as dense vectors that capture their semantic meaning.

\subsection{Generation Phase}

The generation phase comprises two components: the \textit{prompt generator} and the \textit{frozen LLM}. 
A straightforward implementation of the \textit{prompt generator} is to concatenate the documents and the query in a simple sequence, represented as $<D_1, D_2, \dots, D_k, P>$.

Continuing the previous example, the prompt generator constructs the input by combining the retrieved context with the query. A simplified prompt might look like:
\begin{quote} \small 
\textit{\textbf{Instruction:} You are a helpful assistant, given some potential documents and a question, you should first read the potential documents... \newline
\textbf{Documents:} [1] Pleiospilos is a genus of succulent flowering plants... [2] Pueraria is a genus of plants native to Asia...\newline
\textbf{Question:} Which is a flowering plant, Pueraria or Pleiospilos?\newline
\textbf{Answer:}} \end{quote}
The \textit{frozen LLM} then processes this context to generate the correct answer: \textit{``Pleiospilos''}.

Additionally, more sophisticated strategies may include applying various prompt techniques~\cite{CoT, chain-of-note, self-refine, least-to-most} to better align the prompt with the LLM's capabilities, or using iterative methods that involve multiple rounds of retrieval and generation~\cite{ir-cot, flare}. 
The \textit{frozen LLM} is an off-the-shelf LLM used without modification; therefore, this paper does not focus on it.




\section{Empirical Setup} 
\label{sec:exp_setup}

This section presents the setup of our empirical study, including datasets, evaluation metrics, model selection, retrieval configurations, and generation settings.

\subsection{Datasets and Metrics}

\renewcommand{\arraystretch}{1.5}
\renewcommand{\aboverulesep}{0pt}
\renewcommand{\belowrulesep}{0pt}
\begin{table*}
\small
\centering
\caption{Statistics of datasets.}
\label{tab::ExperimentsSettings}
\begin{tabular}{l|ccc|ccc}
\toprule
\textbf{Domain} & \multicolumn{3}{c|}{Software Engineering}  & \multicolumn{3}{c}{Natural Language Processing}  \\ \midrule
\textbf{Dataset} & \textbf{CoNaLa~\cite{conala}} & \textbf{DS1000~\cite{DS1000}} & \textbf{PNE~\cite{pandasEval}} & \textbf{NQ~\cite{NQ}} & \textbf{TriviaQA~\cite{joshi2017triviaqa}} & \textbf{HotpotQA~\cite{yang2018hotpotqa}}  \\
\textbf{Task} & Code Generation & Code Completion & Code Completion & Open-Domain QA & Open-Domain QA & Multihop QA  \\
\textbf{Samples} & 100 & 200 & 200 & 2,000 & 2,000 & 2,000 \\

\bottomrule

\end{tabular}
\end{table*}

Table~\ref{tab::ExperimentsSettings} summarizes our experimental setup across six representative datasets covering scenarios in both software engineering (SE) and natural language processing (NLP) domains. We select representative tasks from both domains to demonstrate diverse challenges and establish generalizable findings across these research areas.

\subsubsection{Software Engineering Tasks.}

We focus on three Python library-oriented datasets: CoNaLa~\cite{conala}, DS1000~\cite{DS1000} and Pandas\_Numpy\_Eval (PNE)~\cite{pandasEval}. These datasets are particularly suitable for RAG evaluation because the core function of samples in those datasets is accomplished by using one or more API functions, making API documentation the primary source of retrieved documents~\cite{wang2024coderag, zhou2022docprompting, liu2023codegen4libs}, enabling well-defined golden (ground-truth) document identification. 
We deliberately exclude open-domain code exercises and repository-level code completion datasets due to their reliance on different retrieval paradigms. Open-domain code exercises typically require retrieval of similar code snippets~\cite{SimilarCodeRetrieval1, SimilarCodeRetrieval2, li2024acecoder}, while repository-level tasks depend on project-specific contextual information~\cite{bairi2024codeplanRepoLevel, zhang2023repocoder, shrivastava2023repositoryLevel, tang2023domainRepoLevel}. Currently, no robust methodologies exist for defining and obtaining golden (ground-truth) documents for these task categories, which would compromise controlled experimental design.

Our empirical study setup includes 200 randomly selected samples from DS1000, all 200 samples from PNE, and all 100 unit-tested samples from CoNaLa~\cite{zhou2022docprompting}. These automatic code generation and completion tasks represent fundamental capabilities for evaluating code-oriented large language models~\cite{lu2021codexglue, humanEval, touvron2023llama, CodeLlama, ChatGPT}.

We evaluate RAG system performance using the \textbf{\textit{Pass@1}} metric, which measures functional correctness by determining whether generated code produces expected outputs when executed with given test inputs. This metric provides robust assessment of code generation quality and is widely adopted in the literature~\cite{humanEval}.

\subsubsection{Natural Language Processing Tasks.}

Question answering (QA) tasks frequently benefit from RAG techniques due to their demand for precise factual knowledge. We select the three most prominent QA datasets: Natural Questions (NQ)~\cite{NQ}, TriviaQA~\cite{joshi2017triviaqa}, and the multi-hop QA dataset HotpotQA~\cite{yang2018hotpotqa}. From each dataset, we randomly extract 2,000 queries to ensure statistical significance while maintaining computational feasibility.

Following established evaluation protocols~\cite{lostinthemiddle}, we assess performance using \textbf{\textit{Accuracy}}, which considers a prediction correct if any of the canonical answers appear as a substring within the generated output.
This approach addresses the limitations of exact string matching by accommodating flexible answer formats while preserving evaluation reliability, as language model outputs may express correct answers through various linguistic formulations.
The robustness of our approach is ensured through the following three aspects:
\begin{enumerate}
        \item \textbf{Constrained Generation}: We utilize strict prompting instructions to force the LLM to generate a single, concise response, thereby preventing the generation of ambiguous multiple-choice lists.
        \item \textbf{Text Normalization}: Before matching, both the model output and ground truth undergo standard normalization, including lowercasing, punctuation removal, and article trimming (a, an, the).
        \item \textbf{Human Verification}: We conducted a human audit on 100 random samples from the NQ dataset. Two annotators verified the outputs and achieved a \textbf{100\%} agreement rate with the automated metric, confirming that the regex-based matching accurately reflects true performance without score inflation.
\end{enumerate}

\subsubsection{Model Selection.}

To ensure comprehensive evaluation and demonstrate the broad applicability of our findings, we deliberately select models spanning different generations, performance levels, and access paradigms. Our model suite includes both closed-source models (GPT-4o-mini and GPT-3.5-turbo from the GPT family~\cite{ChatGPT}) and open-source models (Llama2-13B from the Llama family~\cite{touvron2023llama}). This selection encompasses older, less performant models (Llama2-13B) alongside newer, more capable models (GPT-4o-mini), enabling us to validate whether our conclusions hold across diverse model capabilities.

\subsection{Retrieval Phase}

\subsubsection{Code Corpus Preparation}
\label{subsubsec: code corpus}
Existing Python API documentation collections~\cite{zhou2022docprompting} present several limitations that render them unsuitable for our evaluation: (1) limited coverage of third-party libraries, notably excluding essential packages such as \textit{SciPy}; (2) absence of version-specific information, despite significant API changes between versions (e.g., TensorFlow 1.0 vs. 2.0); and (3) reliance on string matching for retrieval success determination, which produces false positives for APIs sharing identical names (e.g., \textit{tensorflow.keras.layers.add} vs. \textit{tensorflow.python.ops.math\_ops.add}).

To address these limitations, we developed an automated documentation collection and API usage identification pipeline comprising four stages:

\begin{enumerate}
\item \textit{\textbf{API Signature Collection}}: Given a Python library name and version, we programmatically install the library and traverse its module hierarchy to identify all public, callable attributes and extract their API signatures.

\item \textit{\textbf{Document Collection}}: We employ \textit{Pydoc} to gather documentation for each API signature, subsequently merging duplicates where multiple signatures reference identical documentation.

\item \textit{\textbf{Golden API Collection}}: For each code generation task, we parse the Abstract Syntax Tree (AST) of both the unit test program and the canonical solution to extract all function names utilized in the reference implementation.

\item \textit{\textbf{Golden Document Identification}}: We systematically truncate prefixes from each function name in the canonical solution and augment unit tests with \textit{Pydoc} help documentation for each prefix-function combination. API signatures are obtained through execution and matched against the collected documentation to identify ground-truth documents.
\end{enumerate}

We applied this automated pipeline to gather library requirements from all three code datasets, collect comprehensive API documentation, and identify golden documents. Samples without API utilization were excluded from evaluation. Ten percent of samples serve as exemplars for few-shot learning scenarios.
This process yielded a corpus containing 70,956 unique API documents, deduplicated from an initial collection of 365,691 documents.

Given that certain Python API documentations exceed 10,000 tokens and would surpass typical LLM context window limits, we follow the approach established by~\cite{wang2024coderag} and truncate each document to a maximum of 500 tokens to ensure compatibility across all evaluated models while maintaining balanced performance.

\subsubsection{QA Corpus Preparation} 

For the Natural Questions (NQ) and TriviaQA datasets, we utilize the corpus constructed by~\cite{DPR}, which is derived from the English Wikipedia dump dated December 20, 2018. The preprocessing pipeline removes semi-structured data to extract clean textual content from articles, with each article subsequently segmented into disjoint 100-word passages. This process yields a total of 21,015,324 passages for retrieval.
For HotpotQA, we employ the accompanying corpus provided with the original dataset~\cite{yang2018hotpotqa}, comprising 5,233,235 Wikipedia-sourced documents.

Golden document identification follows established methodologies: we adopt the approach from~\cite{DPR} for NQ and TriviaQA datasets, while for HotpotQA, golden documents are determined by verifying alignment with the ground-truth supporting facts provided in the dataset annotations.

\subsubsection{Retriever Design}

To better reflect real-world industrial deployment scenarios, we employ the commercial embedding model \textsc{text-embedding-3-small} from OpenAI as our retriever. This model represents a widely adopted choice in production RAG systems~\cite{OpenAIEmbedding, OpenAIEmbeddingIndustry}, making it a representative choice for evaluating practical deployment scenarios.
This choice prioritizes practical relevance over commonly used experimental baselines, as commercial embedding services represent the predominant approach in production RAG systems.

We use naive retrieving methods, employing the problem description as the retrieval query for our construction approach. After obtaining the embedding vectors of corpus and queries using \textsc{text-embedding-3-small}, we utilize the Faiss~\cite{faiss} library to calculate similarity scores between them.
Since the number of retrieved documents constitutes one of the design choices we analyze (RQ2: How much information should be retrieved), and we need to deliberately control the number of documents to establish controlled experiments, the number of retrieved documents varies across experiments. We introduce specific document counts later in the experimental designs for each research question.

\subsection{Generation Phase}

\subsubsection{Inference Settings.}
To mitigate output variability and ensure reproducible results, we apply greedy decoding (temperature=0) across all evaluated models. This deterministic approach eliminates sampling-related randomness and enables consistent experimental comparisons. Answer extraction from LLM responses is performed using regular expressions tailored to each task format.

We ensure that prompts do not exceed the context window for each LLM in our experimental settings. For QA tasks, we use Llama2-13B with its 4,096-token context window. However, code generation tasks require longer contexts due to extensive API documentation, so we substitute CodeLlama-13B (Llama2-13B fine-tuned on Python code) with its extended 16,384-token context window for all software engineering tasks. This substitution maintains model family consistency while accommodating the longer contextual requirements of code generation.

All inference experiments were conducted on a server equipped with a 4.5GHz AMD 5955WX 16-Core CPU, 256GB RAM, and dual NVIDIA A6000 GPUs (48GB VRAM each). The comprehensive evaluation required over 200,000 API calls for closed-source models and exceeded 2,000 GPU hours for open-source model inference.

\subsubsection{Prompt Generator}
\label{subsubsec:PromptGenerator}
We employ zero-shot instruction prompting as the default approach for integrating retrieved documents with user queries, serving as our baseline methodology. This approach ensures that LLMs accurately comprehend task requirements and generate clearly extractable responses.

Following the classical RAG design established by Lewis et al.~\cite{RAG},
we develop dataset-specific instructions $I$ tailored to each dataset. 
The prompt generator follows the template $\langle I, D\_1, D\_2, \ldots, D\_k, Q \rangle$, where $I$ represents the task-specific instructions, $\langle D\_1, D\_2, \ldots, D\_k \rangle$ denote the $k$ retrieved documents, and $Q$ represents the input query. This template structure mirrors the foundational RAG approach where retrieved documents are systematically integrated into the generation context to enhance factual accuracy and reduce hallucination.

\section{RQ1: Should RAG be deployed?}

\label{sec:RQ1}

To answer this fundamental engineering question, we conduct a systematic evaluation of RAG's value proposition under varying retrieval quality conditions. This approach enables us to determine the conditions under which RAG provides measurable benefits and assess its robustness to retrieval quality degradation.

\subsection{Experimental Design}

We evaluate RAG performance by systematically controlling retrieval recall to determine when RAG outperforms standalone LLMs across various scenarios. We test six discrete recall levels, starting with perfect retrieval (100\% recall, where all golden documents containing relevant answers are retrieved) and progressively degrading performance to identify the threshold at which RAG ceases to provide benefits:

\begin{equation}
Retrieval\ Recall \in \{100\%, 80\%, 60\%, 40\%, 20\%, 0\% \} 
\end{equation}

Our experimental setup recognizes that any RAG system retrieves a mixture of two document types:
\begin{itemize}
\item \textbf{Golden Documents}: Documents that contain authentic answers to the queries, identified as described in Section~\ref{sec:exp_setup}. These represent the ideal documents that a retrieval system should return.
\item \textbf{Distracting Documents}: Documents retrieved by the system that exhibit high semantic similarity to the query but do not contain the correct answer. These documents are topically relevant but ultimately unhelpful for answering the specific question.
\end{itemize} 

We begin with all golden documents and systematically replace them with distracting documents retrieved by the system until reaching each target recall level. At each recall level, we compare RAG performance against standalone LLM performance across different datasets and scenarios to assess when RAG provides measurable advantages.

This controlled recall degradation methodology provides a rigorous framework for answering ``Should I use RAG?'' by empirically establishing when RAG outperforms standalone LLMs and providing practical deployment guidelines.

Our approach is justified for two critical reasons:
\begin{itemize}
\item \textbf{Recall as Dominant Performance Factor}: Retrieval recall represents the most significant factor determining RAG performance~\cite{RAG, RAG_survey}. By systematically controlling this primary variable while testing across diverse scenarios, we can establish fundamental performance boundaries that inform RAG adoption decisions across different domains and use cases.

\item \textbf{Conservative Upper-Bound Analysis}: Our experimental conditions represent optimistic scenarios with clean golden documents, controlled document ratios, and no additional real-world complications (contradictory information, system latency, incomplete knowledge bases). Since other factors like precision and ranking quality will be worse in practice, our recall thresholds represent upper-bound conditions. If RAG cannot outperform standalone LLMs at certain recall levels under these favorable conditions, it will not succeed in more challenging real-world deployments~\cite{RAG_survey, PowerofNoise, lostinthemiddle}.
\end{itemize}

By testing across various popular RAG scenarios, this approach establishes when RAG provides measurable value over standalone LLMs and provides comprehensive insights for practical RAG adoption decisions.

To validate our findings under conditions where retrieval metrics co-vary as they do in production systems, we further conduct a robustness check with retrieval degradation on GPT-4o-mini. While the primary experiment isolates recall to establish upper-bound performance, this condition simulates a realistic, noisy retrieval environment. Specifically, we augment the controlled recall setup by simultaneously applying two perturbations: (1) \textbf{reduced precision}, achieved by injecting five additional topically-relevant distractor documents into the context, and (2) \textbf{degraded ranking}, implemented by randomizing the positions of golden documents within the retrieved set. This configuration allows us to assess whether the performance thresholds established in the controlled analysis remain consistent when retrieval noise and ordering effects are introduced.

\subsection{Results and Findings}

Figure~\ref{fig:ret_recall} summarizes RAG system performance under different retrieval recall conditions across all evaluated datasets and models. The solid lines track RAG system performance while dashed lines represent the baseline LLM performance without retrieval. As shown in the figure, performance generally trends upward as retrieval recall increases. However, the rate of improvement differs by domain: Question Answering (QA) tasks exhibit a steep performance rise (up to 60\%) as recall increases, whereas code generation tasks show a more gradual and constrained improvement (typically 10–20\%). Table~\ref{tab:RetrievalRequirement} details the specific ``threshold recall'' values—the intersection points where RAG performance surpasses the base LLM. These threshold values vary across experiments, showing that different models and tasks require different levels of retrieval quality to exceed the non-retrieval baseline.

Additionally, to ensure these patterns hold under real-world conditions where precision and ranking degrade, we present a robustness check in Table~\ref{tab:robustness_check}, comparing idealized performance against a realistic, noisy retrieval setup.

\subsubsection{Finding-1: RAG applicability highly depends on tasks and models.} \hfill \\

\medskip

\textbf{The threshold retrieval recall varies drastically for different question answering datasets.}
For question answering tasks, the threshold recall values shown in Table~\ref{tab:RetrievalRequirement} reveal significant variation in retrieval recall thresholds required for RAG systems to outperform corresponding base LLMs, ranging from the lowest (0.2) to highest (1.0) values across different scenarios.

Examining the detailed scenarios, NQ and HotpotQA demonstrate consistently low retrieval recall thresholds (0.2 to 0.6), indicating that RAG systems provide benefits when retrieving only a portion of relevant documents. In contrast, TriviaQA exhibits significantly higher thresholds across all models, with most requiring near-perfect retrieval (0.8 to 1.0) to outperform base LLMs. 
This pattern aligns with TriviaQA's question characteristics, consisting primarily of factual questions about well-known entities (e.g., ``Who won Super Bowl XX?''). For such widely-known knowledge that models are already familiar with, RAG provides limited marginal benefits unless retrieval quality is exceptionally high, as models have low probability of knowledge gaps or hallucination on these topics.

\begin{figure*}[ht]
    \centering
    \includegraphics[width=0.9\textwidth]{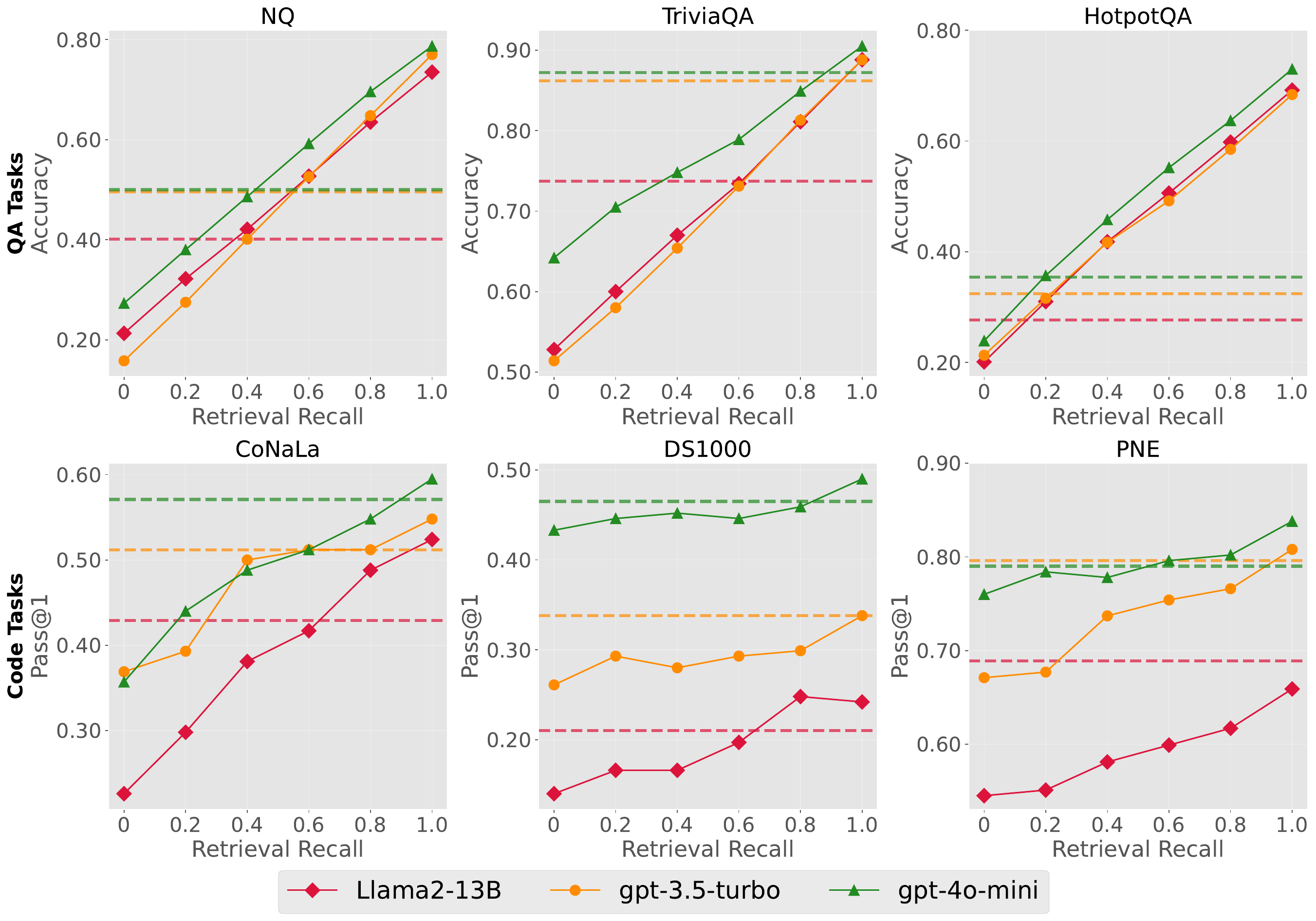}
    \caption{RAG system performance across varying retrieval recall levels and base LLMs. Solid lines show RAG system performance; dashed lines show performance of base LLMs without retrieval.}
    \vspace{-10pt}
    \label{fig:ret_recall}
\end{figure*}

Additionally, GPT-3.5-turbo demonstrates similar performance to Llama2-13B when provided with identical retrieved documents (orange and red line in each subplot in Figure~\ref{fig:ret_recall}), suggesting comparable document comprehension capabilities, while GPT-4o-mini consistently exhibits superior performance across all retrieval conditions (green line in each subplot in Figure~\ref{fig:ret_recall}).

\begin{center}
\begin{table}[t]
    \centering
    \captionsetup{width=0.9\textwidth}
    \small
    \caption{Retrieval recall thresholds across models and datasets. These thresholds represent the minimum retrieval recall among our tested levels needed for RAG systems to outperform the corresponding base LLMs.
    }
    \begin{tabular}{lcccccc}
        \toprule
        \textbf{Model} & \textbf{NQ} & \textbf{TriviaQA} & \textbf{HotpotQA} & \textbf{CoNaLa} & \textbf{DS1000} & \textbf{PNE} \\
        \midrule
        \textbf{Llama-13B} & 0.4   & 0.8 & 0.2 & 0.8 & 0.8 & NA \\
        \textbf{GPT-3.5-turbo} & 0.6 & 1.0 & 0.4 & 0.6 & 1.0 & 1.0 \\
        \textbf{GPT-4o-mini} & 0.6   & 1.0 & 0.4 & 1.0 & 1.0 & 0.6 \\

        \bottomrule
    \end{tabular}
    \label{tab:RetrievalRequirement}
\end{table}
\vspace{-10pt}
\end{center}


\medskip
\textbf{API documentation provides continuous but limited benefits to code generation.}
As shown in the lower three subplots in Figure~\ref{fig:ret_recall}, code generation tasks demonstrate a distinctly different pattern from question answering tasks. 
Performance gains increase smoothly and continuously with retrieval recall, indicating that relevant API documents benefit LLM code generation while irrelevant documents can mislead the generation process.
This result aligns with established findings~\cite{lostinthemiddle, PowerofNoise} that topically similar distractors pose a greater challenge than random noise by interfering with the model's relevance filtering. Consequently, the performance constraints observed here reflect the realistic difficulty of filtering natural retrieval noise, distinct from the artificial difficulty of adversarial attacks.
However, the overall benefits remain constrained—code generation tasks show modest improvements of 10\% (DS1000), less than 20\% (PNE), and just over 20\% (Conala) when moving from recall=0 to recall=100\%, compared to the substantial 60\% improvement observed in question answering tasks like NQ under the same conditions.

Unlike question answering tasks where models can directly extract answers from documents, code generation requires synthesizing retrieved API information into correct, executable code rather than simply locating relevant references.
A comparative example illustrates this distinction in Figure~\ref{fig:APIDocExample}: in NQ, the LLM only needs to find the description ``An ushanka is a Russian fur cap'' in the document and extract the answer ``An ushanka''—document quality directly determines performance. 
However, in PNE, the LLM must understand and properly utilize API documentations from methods like `dropna`, `apply` and `isna`, and apply appropriate operations: `new\_df = df.apply(lambda x: sorted(x, key=pd.isnull)).dropna(how='all')`. Providing proper API documentation assists but does not guarantee successful code generation.

Furthermore, since LLMs already possess substantial knowledge about commonly used library APIs (pandas, numpy, etc.), RAG benefits for code generation appear limited in current evaluation scenarios involving well-known libraries. However, these results establish important baseline expectations for RAG systems when working with private libraries, proprietary APIs, or newly released packages where LLMs lack pre-existing knowledge.
In such scenarios, the knowledge gap makes external documentation retrieval much more valuable for successful code generation, suggesting that API documentation retrieval becomes particularly beneficial when LLMs are unfamiliar with the required APIs.

\definecolor{codegreen}{rgb}{0,0.6,0}
\definecolor{codegray}{rgb}{0.5,0.5,0.5}
\definecolor{codepurple}{rgb}{0.58,0,0.82}
\definecolor{backcolour}{rgb}{0.95,0.95,0.92}
\definecolor{questionbg}{rgb}{0.9,0.9,0.9}
\definecolor{docbg}{rgb}{0.95,0.95,1}
\definecolor{answerbg}{rgb}{0.97,1,0.97}
\definecolor{wronganswerbg}{rgb}{1,0.97,0.97}

\lstdefinestyle{pythonstyle}{
    backgroundcolor=\color{backcolour},   
    commentstyle=\color{codegreen},
    keywordstyle=\color{blue},
    basicstyle=\ttfamily\small,
    breaklines=true,                 
    showspaces=false,                
    showstringspaces=false,
    showtabs=false,                  
    tabsize=2,
    numbers=left,
    numbersep=5pt,
}

\begin{figure}
\centering
\begin{tikzpicture}[scale=0.9, transform shape]

\node[draw, rounded corners=10pt, fill=questionbg, text width=6cm, align=left] (qa_question) at (-4,-1.6) {
\small \textbf{Question:}
What is the name of the Russian fur hat?
};

\node[draw, rounded corners=5pt, fill=docbg, text width=6cm, align=left] (qa_doc) at (-4,-2.7) {
\small \textbf{Document:} "Ushanka \textcolor{red}{An ushanka} (, literally "ear flap hat"), also called a "ushanka-hat", is a Russian fur cap with ear flaps that..."
};

\node[draw, rounded corners=10pt, fill=answerbg, text width=6cm, align=left] (qa_answer) at (-4,-3.6) {
\small \textbf{Answer:} An ushanka
};

\node[draw, rounded corners=10pt, fill=questionbg, text width=8.5cm, align=left] (code_question) at (4,0) {
\small \textbf{Incomplete Code:}
\small \begin{lstlisting}[language=Python, numbers=none, basicstyle=\ttfamily]
import pandas as pd
import numpy as np
df = pd.DataFrame({'A': [1, 4, 7, np.nan], 
                'B': [np.nan, 2, 5, np.nan], 
                'C': [np.nan, np.nan, 3, 6]})
# use sorted to align non NULL data at top, 
# use dropna to drop all rows with all NaN
\end{lstlisting}
};

\node[draw, rounded corners=5pt, fill=docbg, text width=8.5cm, align=left] (code_docs) at (4,-3.3) {
\small \textbf{Documents:}
\small \begin{itemize}
\item \texttt{\textbf{dropna}(axis: 'Axis' = 0, how: 'str' = 'any', thresh=None, subset=None, inplace: 'bool' = False)
Remove missing values. See the ...} 
\item \texttt{\textbf{apply}(func: 'AggFuncType', axis: 'Axis' = 0, raw: 'bool' = False, result\_type=None, args=(), **kwargs)     Apply a function along an axis of...}
\item \texttt{\textbf{isnull}(obj)     
Detect missing values for an array-like...}
\end{itemize}
};

\node[draw, rounded corners=10pt, fill=answerbg, text width=8.5cm, align=left] (code_answer) at (4,-5.8) {
\small \textbf{Generated Code:}
\small \begin{lstlisting}[language=Python, keywordstyle={}, numbers=none, emph={}, emphstyle=\color{red}, basicstyle=\ttfamily]
new_df = df.apply(lambda x:
            sorted(x, key=pd.isnull))
            .dropna(how='all')
\end{lstlisting}
};

\node[align=center] at (-4,-4.2) {\textbf{QA: Direct Extraction}};
\node[align=center] at (4,-7) {\textbf{Code: Synthesis Required}};

\end{tikzpicture}
\caption{How external documents benefit QA versus Code tasks differently. Left: NQ dataset example where documents provide direct factual answers. Right: PNE dataset example where documents provide contextual knowledge for code reasoning.
}
\label{fig:APIDocExample}
\vspace{-10pt}
\end{figure}

\medskip
\textbf{The threshold recall varies for different models.}
Examining model differences in Table~\ref{tab:RetrievalRequirement}, we observe that RAG system with more advanced models like GPT-4o-mini generally requires higher retrieval recall to outperform base LLMs, while Llama2-13B demonstrates lower threshold requirements across all tasks.
Code generation tasks exhibit more complex and variable patterns across different models. For CoNaLa dataset, RAG system with GPT-3.5-turbo needs 60\% retrieval recall to outperform the base LLM, while GPT-4o-mini requires perfect retrieval recall (100\%). Conversely, for PNE, the pattern reverses—GPT-3.5-turbo requires 100\% retrieval recall while GPT-4o-mini needs only 60\%. No clear universal trend governs model-specific retrieval requirements in code generation.

These findings indicate that RAG applicability is closely tied to model selection. Given the same retrieval recall level, different models gain varying levels of benefit, highlighting the need for model-specific optimization strategies rather than universal RAG configurations.

\subsubsection{Finding-2: Golden documents harm RAG system performance in some scenarios.} \hfill \\

\textbf{A substantial proportion of cases are correctly solved by base LLMs but fail when using RAG with golden documents.}
Notably, in the PNE dataset, RAG systems perform worse than base LLMs even when provided with 100\% golden documents. 
To investigate this phenomenon, we conduct a complementary analysis examining prediction patterns between RAG and base LLMs by counting samples that base LLMs correctly solve but RAG systems with golden documents fail. The results are shown in  Table~\ref{tab:OracleVSsingleLLM}. 

The results reveal significant numbers of ``base LLM only correct'' samples across all datasets.
For code generation tasks, up to 12.6\% of samples exhibit this pattern, with an average percentage exceeding 6\%.
For QA tasks, a considerable proportion (2.4\% to 6.9\%) of cases can be answered by base LLMs but fail with RAG systems using golden documents. Even the lowest rate of 2.4\% translates to approximately 50 newly failed cases in large QA datasets (2000 samples).

These failure cases indicate fundamental prediction pattern difference between RAG systems and base LLMs, demonstrating that RAG systems introduce additional failure points that can adversely impact performance.
Analysis of specific failure cases reveals that RAG systems with perfect documents encounter multiple failure points~\cite{SevenFP}, including inability to correctly extract information from retrieved documents (Not Extracted failures) and generation of syntactically incorrect code with undefined variables or missing imports (Wrong Format failures).
Figure~\ref{fig:FailurePoints} illustrates two failure mode cases. The left sub-figure shows an NQ dataset case asking ``when was IISc named to its current name.'' The provided document mentions two key dates: 1958 when the institute was granted university status, and 1909 when it was actually named. The RAG system misinterprets the document context and incorrectly selects 1958 as the answer, while the base LLM correctly identifies 1909 as the naming date.
The right sub-figure shows a CoNaLa dataset case where the task is to get the date 7 days before the current date. When provided with relevant APIs (`now()`, `timedelta()`), the RAG system generates code that uses these functions without properly importing the datetime module, resulting in undefined function errors. In contrast, the base LLM independently generates correct code with proper imports.

These findings suggest that providing relevant documents is necessary but insufficient for RAG success—the system must also effectively utilize retrieved information.
Furthermore, this failure pattern correlates with threshold recall requirements. For instance, GPT-4o-mini on PNE shows only 3\% new failure samples, corresponding to a relatively low threshold (0.6 in Table~\ref{tab:RetrievalRequirement}).
In contrast, Llama2-13B on PNE shows 12.6\% new failed samples, explaining why RAG with perfect golden documents cannot outperform base LLMs-while documents provide benefits in some cases, the system fails substantially in others, resulting in high recall requirements or even under-performance.

\begin{center}
\begin{table*}
    \centering
    \small 
    \caption{Performance distribution analysis between RAG systems and base LLMs}
    \renewcommand{\tabcolsep}{8pt}
    \begin{tabular}{llcccccc}
         \toprule
        LLM & Condition & \textbf{NQ} & \textbf{TriviaQA} & \textbf{HotpotQA} & \textbf{CoNaLa} & \textbf{DS1000} & \textbf{PNE} \\
        \midrule
        \multirow{1}{*}{\textbf{Llama2-13B}}  
        & \textbf{Only correct in base LLM}     & 3.9\%    & 4.3\%       & 2.7\%       & 6.0\%      & 7.0\%    & 12.6\% \\
        
        \midrule
        \multirow{1}{*}{\textbf{GPT-3.5-turbo}} 
        & \textbf{Only correct in base LLM}   & 4.2\%    & 6.9\%       & 2.4\%       & 7.1\%      & 8.9\%      & 7.2\% \\

        \midrule
        \multirow{1}{*}{\textbf{GPT-4o-mini}} 
        & \textbf{Only correct in base LLM}   & 3.7\%    & 4.5\%      & 2.6\%       & 6.0\%      & 5.1\%      & 3.0\% \\ 
        \bottomrule
    \end{tabular}
    \label{tab:OracleVSsingleLLM}
    \vspace{-10pt}
\end{table*}
\end{center}

\begin{figure}[h]
\centering
\begin{tikzpicture}[scale=0.9, transform shape]

\node[draw, rounded corners=10pt, fill=questionbg, text width=6cm, align=left] (qa_question) at (-4,-1.8) {
\small \textbf{Question:}
when was iisc named to its current name
};

\node[draw, rounded corners=5pt, fill=docbg, text width=6cm, align=left] (qa_doc) at (-4,-3.2) {
\small \textbf{Document:} "... In \textcolor{red}{1958} the institute was granted the deemed university status by the UGC. At the time of the inception of IISc in \textcolor{red}{1909}, Morris Travers, Sir William Ramsay's co-worker in the ..."
};

\node[draw, rounded corners=10pt, fill=wronganswerbg, text width=6cm, align=left] (qa_answer) at (-4,-4.5) {
\small \textbf{LLM Response:} 1958
};

\node[draw, rounded corners=10pt, fill=answerbg, text width=6cm, align=left] (qa_answer) at (-4,-5) {
\small \textbf{Correct Answer:} 1909
};

\node[draw, rounded corners=10pt, fill=questionbg, text width=8.5cm, align=left] (code_question) at (4,0) {
\small \textbf{Problem Description:} get the date 7 days before the current date
};

\node[draw, rounded corners=5pt, fill=docbg, text width=8.5cm, align=left] (code_docs) at (4,-1.65) {
\small \textbf{Documents:}
\small \begin{itemize}
\item \texttt{\textbf{now}(tz=None)
    Returns new datetime object representing current...} 
\item \texttt{class \textbf{timedelta}(builtins.object)
Difference between two datetime values.
timedelta(days=0, seconds=0, microseconds=...}
\item \texttt{\textbf{date}()
    Return date object with same year, month...}
\end{itemize}
};

\node[draw, rounded corners=10pt, fill=wronganswerbg, text width=8.6cm, align=left] (code_answer) at (4,-4.15) {
\small \textbf{Generated Code:}
\small \begin{lstlisting}[language=Python, keywordstyle={}, numbers=none, emph={now, timedelta}, emphstyle=\color{red}, basicstyle=\ttfamily]
def f_20573459():
    today = now()
    seven_days_ago = today - timedelta(days=7)
    return seven_days_ago.date()
\end{lstlisting}
};

\node[draw, rounded corners=10pt, fill=answerbg, text width=8.5cm, align=left] (code_answer) at (4,-6.5) {
\small \textbf{Correct Code:}
\small \begin{lstlisting}[language=Python, keywordstyle={}, numbers=none, emph={}, emphstyle=\color{red}, basicstyle=\ttfamily]
import datetime
def f_20573459():
    return (datetime.datetime.now() - 
    datetime.timedelta(days=7)).date()
\end{lstlisting}
};

\node[align=center] at (-4,-5.5) {\textbf{QA: Misunderstand Document}};
\node[align=center] at (4,-8) {\textbf{Code: Use Undefined Function}};

\end{tikzpicture}
\caption{Two failure cases where RAG systems fail despite having access to golden documents, while base LLMs succeed. Left: In NQ dataset sample, RAG system misunderstands the document and returns 1958 instead of the correct answer 1909. Right: In CoNaLa dataset sample, RAG system generates code with undefined functions despite having relevant documentation. 
}
\label{fig:FailurePoints}
\vspace{-10pt}
\end{figure}

\subsubsection{Finding-3: Performance trends and task-specific characteristics remain robust under realistic retrieval degradation.} \hfill \\

\textbf{The performance trajectories persist when retrieval precision and ranking quality degrade.} To address the concern that isolating recall idealizes real-world systems, we conducted a robustness check on GPT-4o-mini under a ``Realistic'' condition (Table~\ref{tab:robustness_check}). This setup simulates production environments by simultaneously introducing retrieval noise (via distractor injection) and degrading ranking quality (via position randomization). As expected, these perturbations cause a general attenuation in absolute performance; however, this degradation manifests primarily as a uniform offset (typically 2--6\% drop in most cases) rather than a disruption of the underlying trend. Because the performance penalty is applied consistently across recall levels, the shape of the performance curves is preserved. Consequently, the distinct patterns observed in Finding-1—such as the steep improvement in QA tasks versus the constrained gains in code generation—remain clearly visible. This structural consistency confirms that the recall thresholds identified in our controlled experiments serve as valid proxies for RAG behavior, even in complex, noisy environments.

\textbf{The reduced performance advantage under realistic conditions further validates the necessity of deployment thresholds.}
The observed performance drop in the realistic setting narrows the margin between the RAG system and the base LLM. In production scenarios where retrieval is noisy, surpassing the non-retrieval baseline becomes significantly more challenging. While a system might appear beneficial under ideal retrieval conditions, realistic noise can easily depress its performance below that of the base LLM. This reality amplifies the importance of our first research question: determining whether to deploy RAG requires strictly verifying that the system can withstand inevitable retrieval degradation without falling below the baseline.

\begin{table*}[t]
\centering
\small
\caption{Robustness check with retrieval degradation on GPT-4o-mini. Comparison of downstream performance between the \textit{Ideal} condition (perfect ranking, isolated recall) and the \textit{Realistic} condition (added distractor noise, randomized ranking).
}
\label{tab:robustness_check}
\setlength{\tabcolsep}{4pt}
\begin{tabular}{l|c|cccccc}
\toprule
\textbf{Dataset} & \textbf{Condition} & \textbf{Recall=0.0} & \textbf{Recall=0.2} & \textbf{Recall=0.4} & \textbf{Recall=0.6} & \textbf{Recall=0.8} & \textbf{Recall=1.0} \\
\midrule
\multirow{2}{*}{\textbf{NQ}} 
& Ideal & 0.273 & 0.380 & 0.486 & 0.592 & 0.696 & 0.787 \\
& Realistic & 0.254 & 0.332 & 0.419 & 0.501 & 0.582 & 0.661 \\
\midrule
\multirow{2}{*}{\textbf{TriviaQA}} 
& Ideal & 0.642 & 0.705 & 0.748 & 0.789 & 0.849 & 0.905 \\
& Realistic & 0.585 & 0.623 & 0.679 & 0.723 & 0.776 & 0.830 \\
\midrule
\multirow{2}{*}{\textbf{HotpotQA}} 
& Ideal & 0.239 & 0.357 & 0.458 & 0.552 & 0.637 & 0.730 \\
& Realistic & 0.257 & 0.349 & 0.443 & 0.525 & 0.615 & 0.704 \\
\midrule
\multirow{2}{*}{\textbf{CoNaLa}} 
& Ideal & 0.357 & 0.440 & 0.488 & 0.512 & 0.548 & 0.595 \\
& Realistic & 0.417 & 0.429 & 0.488 & 0.512 & 0.512 & 0.571 \\
\midrule
\multirow{2}{*}{\textbf{DS1000}} 
& Ideal & 0.433 & 0.446 & 0.452 & 0.446 & 0.459 & 0.490 \\
& Realistic & 0.407 & 0.420 & 0.446 & 0.446 & 0.459 & 0.478 \\
\midrule
\multirow{2}{*}{\textbf{PNE}} 
& Ideal & 0.760 & 0.784 & 0.778 & 0.796 & 0.802 & 0.838 \\
& Realistic & 0.719 & 0.731 & 0.731 & 0.749 & 0.743 & 0.772 \\
\bottomrule
\end{tabular}
\end{table*}

\begin{tcolorbox}
    \textbf{Answer to RQ1:} 
    RAG deployment should be highly selective due to dramatic context variation: retrieval thresholds required for outperforming base LLMs range from 0.2 to 1.0, and performance gains vary from substantial (0.6 accuracy improvement in QA) to limited (0.1-0.2 pass@1 improvement in code generation).
    These margins are further compressed in realistic production environments, where retrieval noise and ranking degradation make surpassing the base LLM significantly more challenging.
    Critically, RAG systems fail on cases that base LLMs solve correctly—affecting up to 12.6\% of samples even with perfect documents—due to document misinterpretation and improper utilization. Therefore, RAG deployment requires careful evaluation of task characteristics, model capabilities, and tolerance for new failure modes rather than universal adoption.
\end{tcolorbox}

\section{RQ2: How much information should be retrieved?} 
\label{sec:RQ2}

\subsection{Experimental Design}

To address this fundamental parameter selection challenge, we employ a two-phase experimental approach that provides both empirical optimization guidelines and practical deployment strategies.

\textbf{Phase 1: Empirical Investigation} systematically varies the number of retrieved documents ($k$) across all datasets and models to establish comprehensive empirical understanding of optimal $k$ selection patterns under various scenarios. This provides essential baseline knowledge of performance-cost trade-offs for different contexts that currently lack systematic investigation.

Increasing $k$ introduces a fundamental trade-off: while more documents may include additional golden documents containing relevant information, they also introduce more distracting documents that can harm performance. We test empirically selected $k$ values ranging from minimal ($k=1$) to maximum practical limits approaching context window constraints.
Given the document length differences between domains (NLP tasks average ~150 tokens vs. SE tasks average ~300 tokens), we adopt different $k$ ranges to maintain comparable context usage. We investigate:
\begin{gather}
    k \in \{1,3,5,10,15,20,30,40\}, Question\ Answering \\
    k \in \{1,3,5,7,10,13,16,20\}, Code\ Generation
\end{gather}

\textbf{Phase 2: Uncertainty-Guided Selection} investigates whether model uncertainty can serve as a practical proxy for $k$ optimization in new, untested scenarios where ground truth evaluation is expensive or unavailable. This addresses the critical deployment challenge where practitioners need $k$ optimization guidance but cannot afford extensive performance testing or lack labeled data for evaluation.

As an early-stage investigation, we analyze whether perplexity—a readily available uncertainty measure from model inference—correlates with the optimal $k$ patterns identified in Phase 1.
Perplexity quantifies model uncertainty about generated text, calculated as:
\begin{equation}
\text{PPL} = \exp\left(-\frac{1}{N} \sum_{i=1}^{N} \log p(x_i)\right).
\end{equation}
where $p(x\_i)$ represents the probability of each generated token.

To ensure robust calculation across different model types, we standardize the probability acquisition and token scoping processes.
For open-source models (e.g., Llama2-13B), token probabilities are derived directly from the output logits. For closed-source models (GPT-3.5-turbo, GPT-4o-mini), we utilize the \texttt{logprobs} parameter in the OpenAI API to retrieve the log-probability of each generated token.
Crucially, to prevent formatting tokens or ``chat'' filler from diluting the metric, we calculate \textit{answer-only perplexity}. Our prompts strictly instruct models to enclose their final output within XML tags (e.g., \texttt{<answer>} or \texttt{<code>}). We parse the full generation to isolate the content within these tags and calculate perplexity exclusively on this solution segment, ignoring all surrounding text.

Phase 1 provides empirical insights for tested scenarios, while Phase 2 enables practitioners to optimize $k$ in new scenarios by testing different $k$ values and using only perplexity measurements (no ground truth required) to predict optimal performance based on established correlation patterns.
This two-phase approach directly answers RQ2 by providing both systematic empirical evidence for $k$ optimization across diverse scenarios and a practical uncertainty-based method for $k$ selection in deployment scenarios where performance evaluation is prohibitively expensive.

\subsection{Results and Findings}

\begin{figure*}
    \centering
    \includegraphics[width=0.9\textwidth]{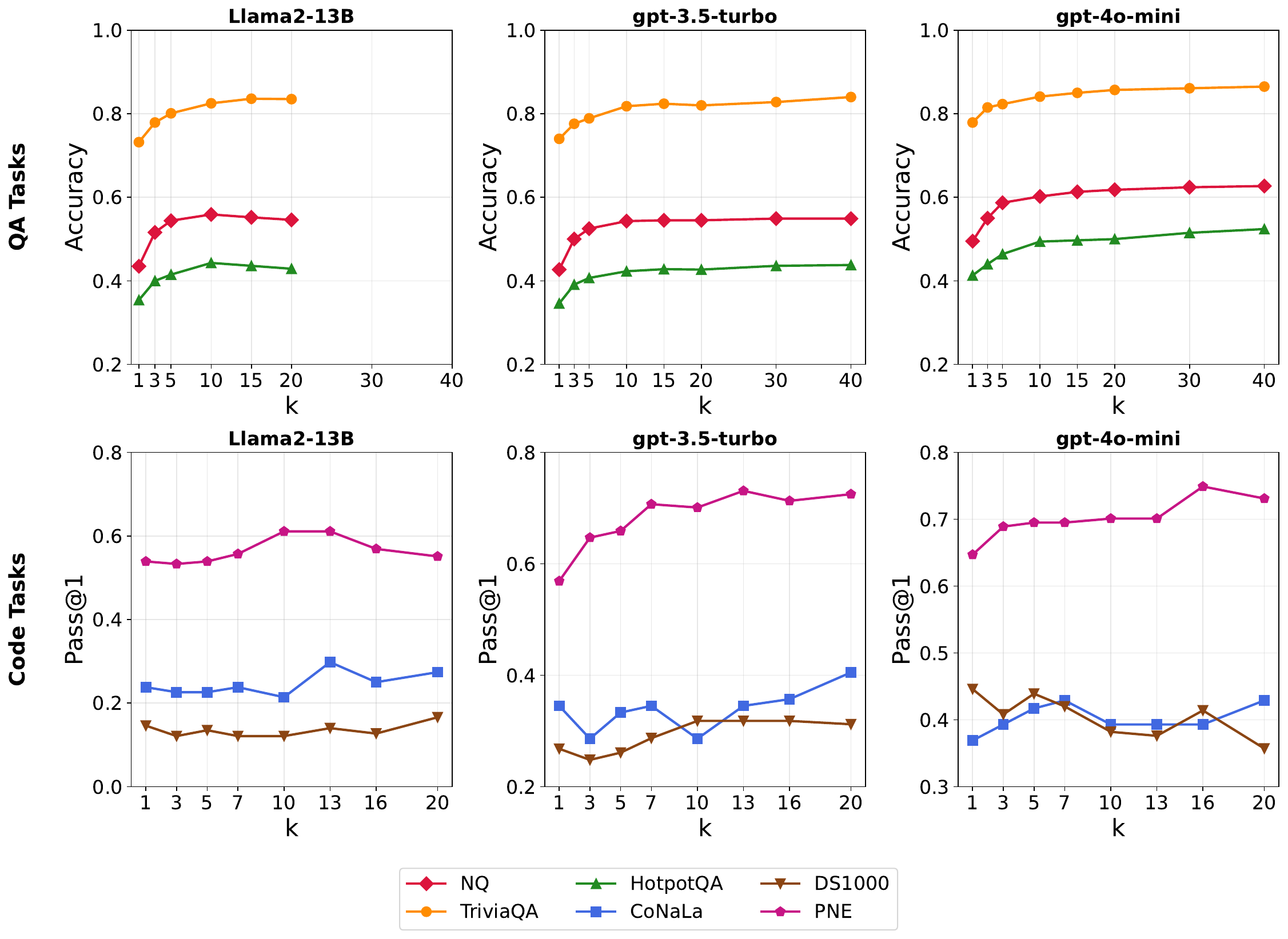}
    \caption{RAG systems performance with varying numbers of retrieved documents. The RAG system using Llama2-13B shows results only up to $k=20$ due to context window constraints. 
    }
    \label{fig:doc_num_analysis}
    \vspace{-15pt}
\end{figure*}

We summarize the performance of RAG systems under varying numbers of documents across various datasets and models in Figure~\ref{fig:doc_num_analysis}. For clarity, each subplot in the figure shows the RAG performance with one LLM across either three question answering tasks or code generation tasks. For Question Answering (QA) tasks, the data shows a consistent trend: performance rises sharply as documents are added, then plateaus or slightly declines. In contrast, code generation tasks exhibit significant variation across different models, with no uniform pattern observed across all scenarios.

\subsubsection{Finding-1: QA tasks share similar optimal document numbers, while code tasks vary significantly.} \hfill \\

\medskip
\textbf{The optimal document number $k$ setting for QA tasks is between 5 and 10.}
As shown in Figure~\ref{fig:doc_num_analysis}, all RAG systems performance in question answering tasks shows a trend of first greatly rising then becoming stable for GPT-4o-mini and GPT-3.5-turbo, or slightly dropping for Llama2-13B due to its weaker information processing ability and context window limitations.
The inflection point that transitions performance from sharp increase to slight increase or even decline occurs at either $k=5$ or $k=10$, as observed in each line in the upper three subplots of Figure~\ref{fig:doc_num_analysis}.

To rigorously validate whether increasing the document count yields significant benefits, we conducted McNemar's tests between consecutive document groups (e.g., $k=1$ vs $k=3$) with Benjamini-Hochberg FDR correction, shown in Table~\ref{tab:mcnemar_results}. The statistical analysis confirms this trend: prior to $k=10$, increasing $k$ yields significant performance improvements in all but one case (Llama2-13B between $k=5$ and $k=10$). Furthermore, comparisons beyond $k=10$ show no significant improvement in the majority of cases. While two statistically significant instances occurred at higher depths ($k=20 \to 30$ and $30 \to 40$), the substantial increase in $k$ requires disproportionate token consumption and latency for only marginal gains.
We also calculated absolute effect sizes and CIs (Appendix Table~\ref{tab:appendix_stats}). Before $k=10$, increasing $k$ typically yields absolute accuracy gains ranging from 0.015 to 0.029, whereas beyond $k=10$, the changes diminish to a range of -0.007 to 0.015. This demonstrates that in most QA scenarios, the optimal $k$ lies between 5 and 10, revealing a consistent pattern.

\begin{figure*}
    \centering
    \includegraphics[width=0.6\textwidth]{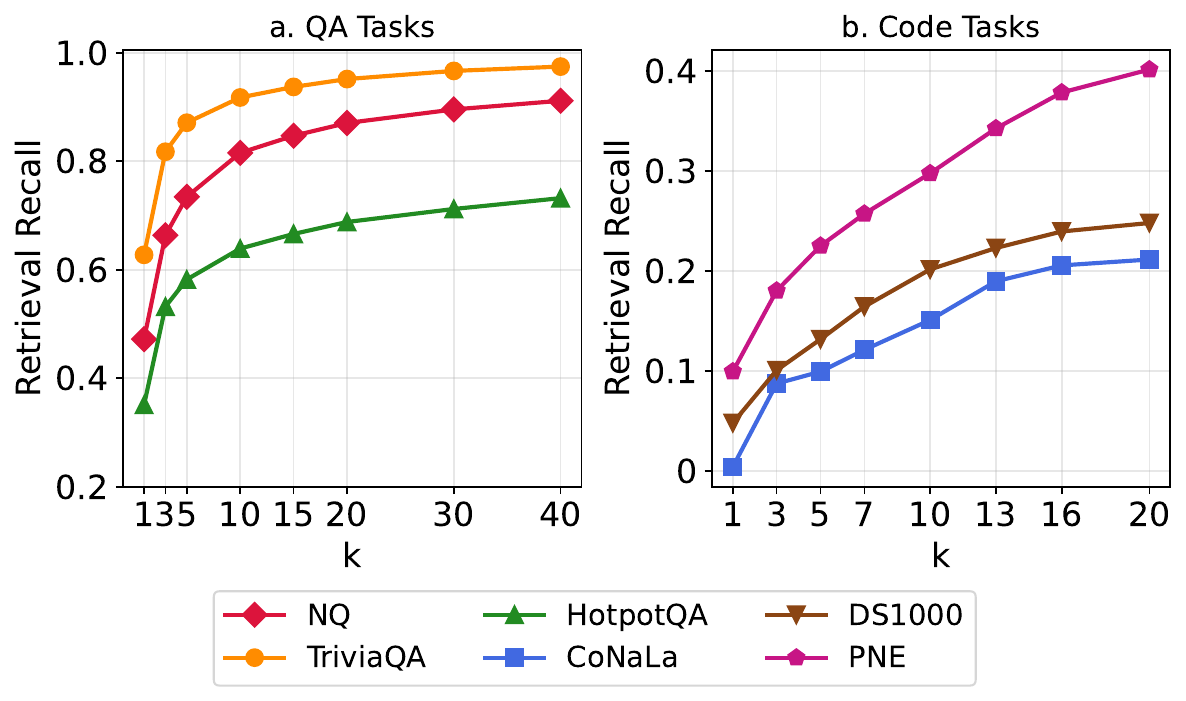}
    \caption{Retrieval recall of RAG systems under varying numbers of documents.}
    \label{fig:ret_recall_for_retriever}
\end{figure*}

This pattern can be explained by the similar retrieval recall trends observed across the three different QA tasks. As shown in Figure~\ref{fig:ret_recall_for_retriever}, the retrieval recall of all three datasets follows a pattern of sharp initial increase, then gradual leveling after $k=10$. This shared retrieval pattern results in the shared optimal $k$ pattern across QA tasks, even for more complex multi-hop reasoning tasks like HotpotQA.

\renewcommand{\H}{\checkmark}             
\newcommand{\M}{\dag}
\newcommand{\N}{\texttimes} 

\begin{table}
\centering
\small
\caption{McNemar’s test results for consecutive document number increases across QA datasets and models. Significance is determined after applying the Benjamini-Hochberg FDR correction. \H\H~indicates highly significant improvement ($p_{adj} < 0.01$), \H~indicates significant improvement ($0.01 \le p_{adj} < 0.05$), \M~indicates \textit{marginally significant} improvement ($0.05 \le p_{adj} < 0.10$), and \N~indicates no significant difference ($p_{adj} \ge 0.10$).
}
\label{tab:mcnemar_results}
\begin{tabular}{|l|ccc|ccc|ccc|}
\hline
\multirow{2}{*}{Comparison} & \multicolumn{3}{c|}{\textbf{Llama2-13B}} & \multicolumn{3}{c|}{\textbf{GPT-3.5-turbo}} & \multicolumn{3}{c|}{\textbf{GPT-4o-mini}} \\
\cline{2-10}
& \textbf{NQ} & \textbf{TriviaQA} & \textbf{HotpotQA} & \textbf{NQ} & \textbf{TriviaQA} & \textbf{HotpotQA} & \textbf{NQ} & \textbf{TriviaQA} & \textbf{HotpotQA} \\
\hline
$k=1$ vs $k=3$ & \H\H   & \H\H & \H\H & \H\H & \H\H & \H\H & \H\H & \H\H & \H\H \\
$k=3$ vs $k=5$ & \H\H   & \H\H & \H\H & \H\H & \M   & \M   & \H\H & \N   & \H\H \\
$k=5$ vs $k=10$ & \N    & \H\H & \H\H & \H   & \H\H & \M   & \H   & \H\H & \H\H \\
$k=10$ vs $k=15$ & \N   & \N   & \N   & \N   & \N   & \N   & \M   & \N   & \N \\
$k=15$ vs $k=20$ & \N   & \N   & \N   & \N   & \N   & \N   & \N   & \N   & \N \\
$k=20$ vs $k=30$ & -    & -    & -    & \N   & \N   & \N   & \N   & \N   & \H \\
$k=30$ vs $k=40$ & -    & -    & -    & \N   & \H   & \N   & \N   & \N   & \N \\
\hline
\end{tabular}
\end{table}

\medskip

\begin{figure*}
    \centering
    \includegraphics[width=0.65\textwidth]{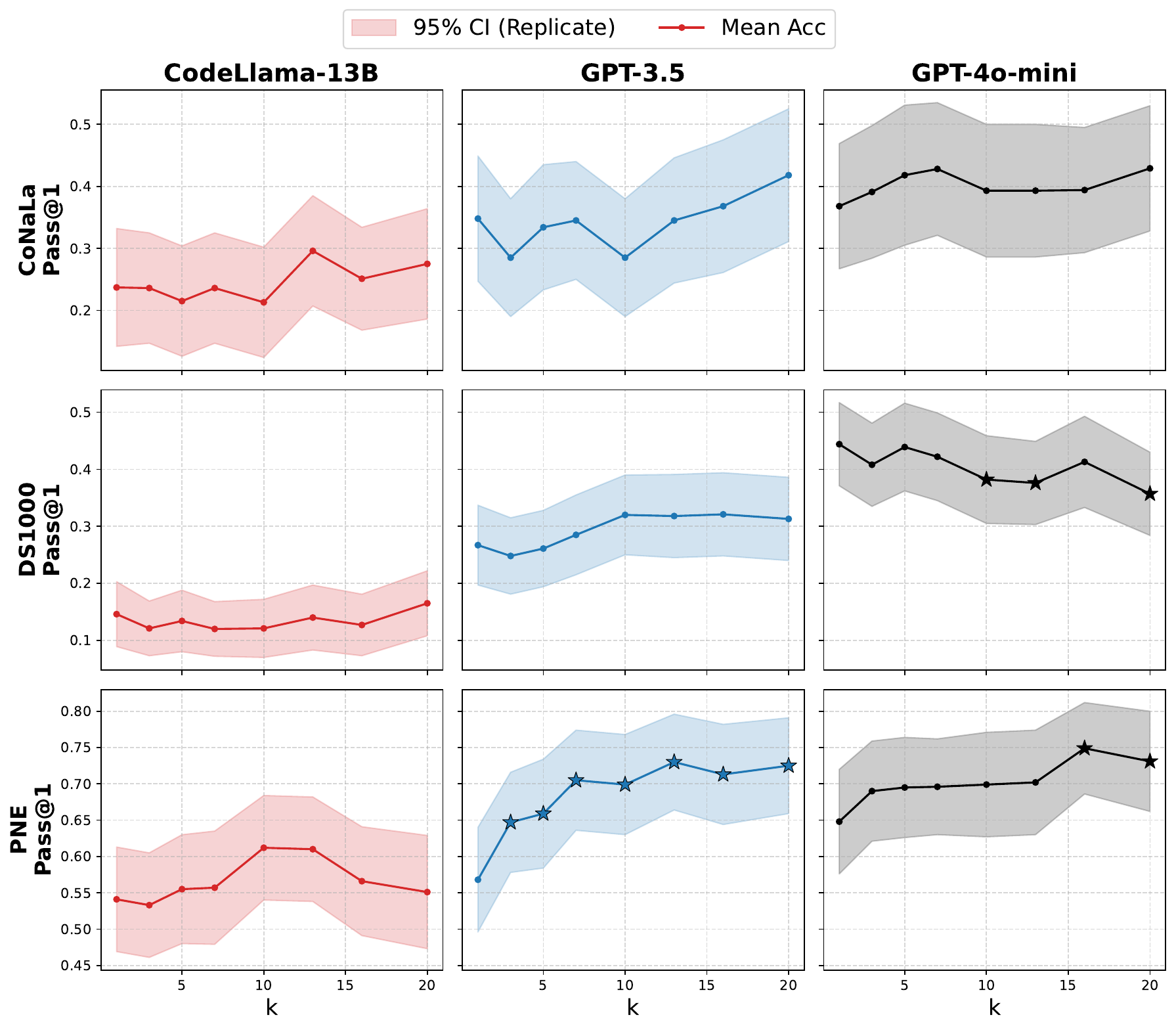}
    \caption{Bootstrap replicate analysis of document number $k$ on the performance of code generation datasets. Solid lines denote the mean \textit{pass@1} score averaged across bootstrap iterations, while shaded regions represent the 95\% confidence interval, visually quantifying generation variance. The symbol $\star$ marks specific scenarios where the performance gain over the baseline $k=1$ is statistically significant ($p<0.05$) based on our paired bootstrap test.}
    \label{fig:code_replicate_folds}
\end{figure*}

\textbf{For code tasks, RAG performance generally exhibits statistical insensitivity to expanded context, primarily driven by insufficient retrieval recall.} Unlike QA scenarios, code generation tasks exhibit highly unpredictable patterns across the three datasets and models. For DS1000 (brown lines), Llama2-13B shows minimal variation close to 20\% Pass@1, while GPT-4o-mini actually declines from 45\% to 35\%. CoNaLa (blue lines) demonstrates even more erratic behavior, characterized by wide error bands (shaded regions in Figure~\ref{fig:code_replicate_folds}) and flat trajectories where performance changes are subtle from variance. Only PNE (purple lines) demonstrates QA-like behavior with consistent improvements.

This instability stems from a fundamental constraint: \textbf{the marginal utility of additional context for code tasks is often too low to overcome generation variance.}
As shown in Figure~\ref{fig:ret_recall_for_retriever}, retrieval recall for datasets like DS1000 and CoNaLa plateaus early and remains significantly lower than typical QA benchmarks.
Consequently, increasing $k$ rarely provides the critical information needed to correct a failed generation, but linearly increases the input noise.
Because the recall gain is functionally flat, the slight benefits of retrieval fail to steer the model towards a correct solution.
Our paired bootstrap analysis in Figure~\ref{fig:code_replicate_folds} statistically confirms this mechanism: for CoNaLa and DS1000, the performance fluctuations across $k$ are not statistically significant.

To determine whether this insensitivity stems from the weak retrieval signal or simply limited sample sizes, PNE can be a comparative case.
PNE shares a comparable sample size with CoNaLa and DS1000 but possesses a slightly stronger retrieval recall increase.
Under identical testing conditions, PNE exhibits a statistically significant positive trend ($p<0.05$, marked by stars) for GPT-3.5-turbo and GPT-4o-mini.
In contrast, DS1000 and CoNaLa show no significant gain despite similar sample sizes.
This dissociation confirms that the statistical insignificance in CoNaLa and DS1000 is largely not due to the limited sample size; rather, it indicates that the retrieval recall improvement brought by the document number increasing is genuinely too weak to drive a robust trend.

However, this does not imply that code generation is inherently incompatible with RAG; rather, it highlights a dependency on retrieval quality.
This is evidenced by the PNE dataset, which maintains a higher recall baseline than CoNaLa and DS1000. 
Because PNE provides a sufficient signal-to-noise ratio, it is the only dataset to exhibit a statistically significant positive trend in our analysis.
This finding offers a clear path forward: while current sparse retrievers often fail to trigger robust scaling, we hypothesize that future dense retrievers capable of higher recall could unlock the stable scaling laws observed in QA and PNE.
Under current conditions, however, increasing $k$ offers no reliable benefit for most code tasks.

Furthermore, we observe clear evidence of information overload, where excessive noise actively harms performance.
On DS1000, GPT-4o-mini exhibits a statistically significant decline in performance as $k$ increases; on PNE, CodeLlama-13B exhibits a ``first rise then drop'' trend, both largely due to the inclusion of noisy information.

\begin{figure*}[ht]
    \centering
    \includegraphics[width=0.9\textwidth]{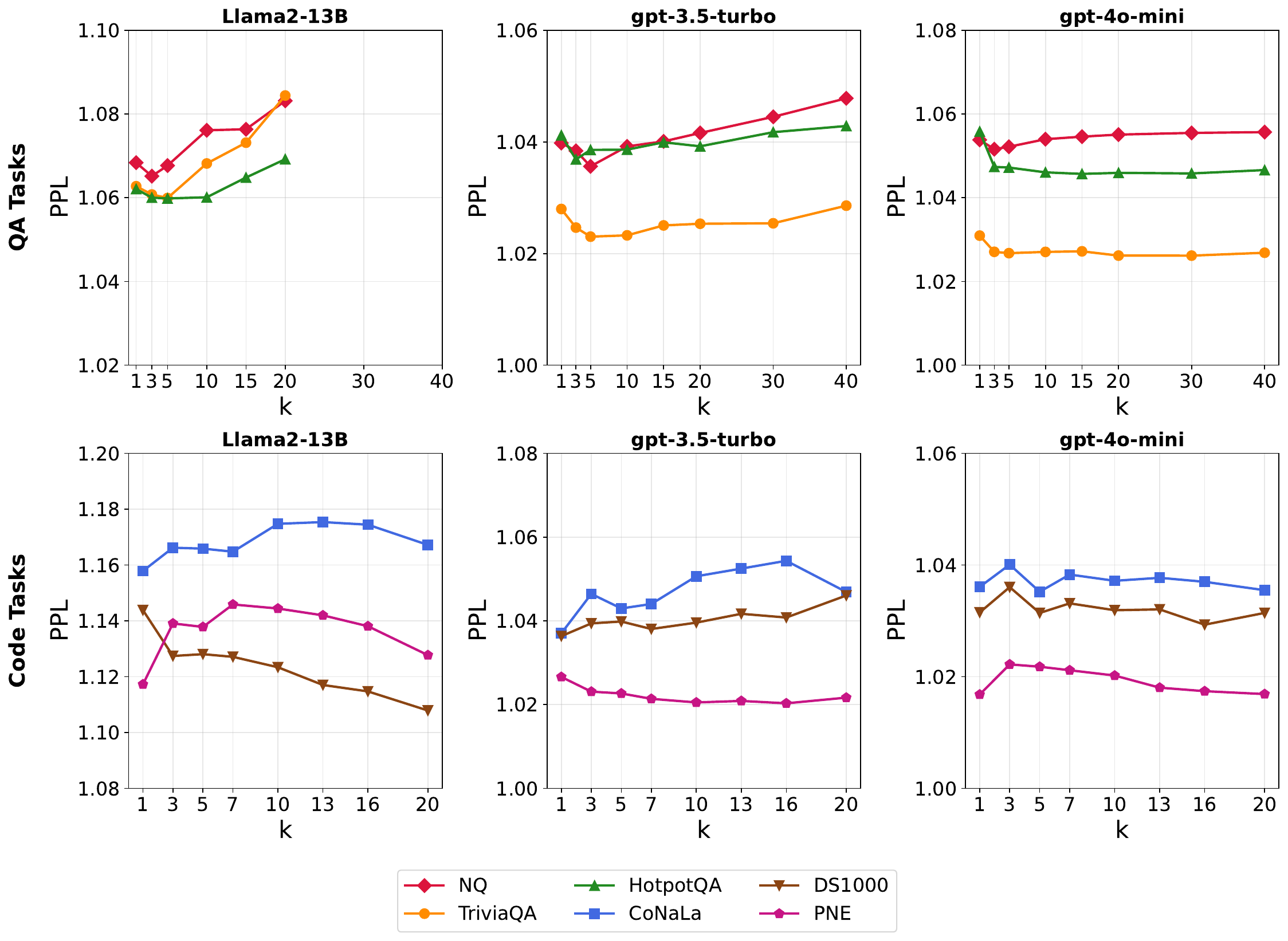}
    \caption{Average of Perplexity in RAG systems generation with varying numbers of retrieved documents. The RAG system using Llama2-13B shows results only up to $k=20$ due to context window constraints.
    }
    \label{fig:doc_num_PPL}
\end{figure*}

\begin{table}[h]
\centering
\small
\caption{Spearman's rank correlation ($\rho$) between perplexity and accuracy. The correlation is calculated across the varying retrieval document counts ($k$) for each model-dataset pair. A positive $\rho$ indicates perplexity rises with accuracy, while a negative $\rho$ indicates perplexity drops as accuracy improves. Significance levels: \H\H~($p<0.01$), \H~($p<0.05$), \M~($p<0.10$).}
\label{tab:ppl_corr}
\setlength{\tabcolsep}{8pt}
\begin{tabular}{l c c c}
\toprule
\textbf{Dataset} & \textbf{Llama2-13B} & \textbf{GPT-3.5-turbo} & \textbf{GPT-4o-mini} \\
\midrule
NQ       & $0.60$   & $0.82^{\H}$  & $0.93^{\H\H}$ \\
TriviaQA & $0.71$   & $0.40$       & $-0.64^{\M}$ \\
HotpotQA & $0.26$   & $0.64^{\M}$  & $-0.69^{\M}$ \\
\midrule
CoNaLa   & $0.27$   & $0.25$       & $-0.13$ \\
DS1000   & $-0.17$  & $0.51$       & $-0.17$ \\
PNE      & $0.54$   & $-0.69^{\M}$ & $-0.29$ \\
\bottomrule
\end{tabular}
\end{table}

\subsubsection{Finding-2: perplexity can serve as a practical proxy for optimal document selection in QA tasks}

Beyond identifying optimal document numbers through performance evaluation, we investigate whether uncertainty metrics can provide practical guidance for real-world k selection without requiring labeled test data.
We examine the relationship between generation perplexity and RAG performance across varying document numbers (Figure~\ref{fig:doc_num_PPL}). Due to setting the temperature to 0 for experimental stability and reproducibility, the perplexity values and variations are quite low.

\medskip
\textbf{For QA tasks, perplexity serves as a useful proxy for optimal document selection, with model-specific patterns}
As observed in Figure~\ref{fig:doc_num_PPL}, perplexity patterns exhibit distinct trajectories that correlate with architectural capabilities. Llama2-13B demonstrates continuous perplexity increase beyond k=5, reflecting its limited 4K context window and weaker ability to synthesize multiple documents—additional documents introduce noise that the model cannot effectively filter. GPT-3.5-turbo shows moderate perplexity increases with its 16K context window, suggesting better but still limited noise tolerance. Most remarkably, GPT-4o-mini's perplexity continues decreasing throughout the entire k range, perfectly mirroring its consistent performance improvements and demonstrating superior document synthesis capabilities with its 128K context window.

An important observation emerges when comparing perplexity minima with performance optima: for Llama2-13B and GPT-3.5-turbo, lowest perplexity occurs at $k=3$ to $5$ while optimal performance appears at $k=5$ to $10$. This offset reveals a fundamental trade-off—while additional documents beyond the perplexity minimum introduce generation uncertainty (reflected in higher perplexity), they simultaneously provide critical complementary information that enhances factual accuracy. However, GPT-4o-mini breaks this pattern entirely, showing both decreasing perplexity and improving performance simultaneously, indicating its ability to extract value from additional documents without uncertainty penalties.

To quantify this relationship, we calculated the Spearman's rank correlation ($\rho$) between answer-only perplexity and accuracy across all $k$ levels (Appendix Table~\ref{tab:ppl_corr}). Theoretically, one would expect a strong negative correlation: as the model becomes more confident (lower PPL), accuracy should increase.
Our data partially supports this: complex reasoning tasks like HotpotQA (with GPT-4o-mini) exhibit this expected negative trend ($\rho = -0.69$), where the model's increasing confidence aligns with better performance.

However, the relationship is not universally linear. For datasets like NQ, we observe a counter-intuitive positive correlation (e.g., $\rho = 0.93$ for GPT-4o-mini). A closer examination of the curves reveals why: Accuracy and Perplexity decouple at higher retrieval depths.
In the NQ case, Perplexity follows a ``U-shaped'' trajectory, reaching a distinct minimum early (at $k=3$) before rising again as additional documents introduce context noise. In contrast, Accuracy continues to rise monotonically (from 0.60 at $k=3$ to 0.62 at $k=40$) because the model is robust enough to extract information despite the noise.
Consequently, although the PPL minimum ($k=3$) correctly identifies the ``efficiency sweet spot'' where performance gains are steepest, the statistical correlation becomes positive because the tail of the curve is dominated by rising PPL and rising Accuracy.

This indicates that while the PPL-Accuracy relationship is not a simple 1-to-1 linear mapping, the PPL minimum serves as a critical indicator of the ``efficiency sweet spot'' (the bottom of the U-shape). Beyond this point, while accuracy may continue to rise slightly, the marginal gains diminish significantly (mirroring our finding that benefits plateau after $k=10$). Thus, PPL effectively signals the point of diminishing returns, acting as a conservative guardrail for resource-efficient retrieval, although we note that fully characterizing this non-monotonic relationship remains a subject for future investigation.

These findings suggest model-tier-specific retrieval strategies. Resource-constrained models like Llama2-13B should prioritize fewer documents ($k=3$ to $5$) to minimize noise-induced uncertainty. Mid-tier models like GPT-3.5-turbo benefit from moderate document counts ($k=5$ to $7$) that balance information gain with manageable uncertainty increases. Most significantly, high-capability models like GPT-4o-mini can effectively process extremely large document sets—when computational latency is not critical, practitioners can retrieve substantially more documents ($k=20$ or more) as these models continue extracting value from additional information without suffering uncertainty penalties. This observation fundamentally challenges the notion of universal optimal k values and suggests that retrieval strategies should scale with model capabilities.

\medskip
\textbf{For code tasks, perplexity shows limited utility as a selection criterion.}
Comparing the perplexity levels and performance in code tasks from Figure~\ref{fig:doc_num_analysis} and \ref{fig:doc_num_PPL} reveals inconsistent correlations across model-dataset combinations. While some alignments exist—GPT-4o-mini on PNE shows lowest perplexity at k=16 matching highest performance, and similar patterns appear on DS1000 at k=5—many cases demonstrate contradictory relationships. For instance, GPT-3.5-turbo on CoNaLa shows performance drops at k=3 corresponding with perplexity rises, yet other data points break this pattern.
This visual inconsistency is statistically corroborated by the Spearman's rank analysis in Table~\ref{tab:ppl_corr}. Unlike QA tasks, code generation shows no statistically significant correlation between perplexity and accuracy in the vast majority of cases (8 out of 9 pairings). The correlation coefficients ($\rho$) fluctuate unpredictably between weak positive (e.g., $0.51$ for DS1000/gpt-3.5) and weak negative values (e.g., $-0.17$ for DS1000/Llama2-13B), confirming the absence of a systematic predictive relationship.

The inconsistency stems from fundamental differences in how weaker models handle code generation with multiple documents. Llama2-13B and GPT-3.5-turbo frequently generate extraneous content (e.g., they repeatedly generate the given API documents before generating the code) or produce outputs unrelated to the coding instructions, making perplexity measurements unreliable for document selection. When models generate off-topic content, perplexity becomes incomparable across different $k$ values, explaining why correlations appear sporadically rather than systematically. This pattern aligns with findings in~\cite{spiess2024calibration}, which demonstrate that token probability does not correlate with code generation correctness for base LLMs.

This contrasts with QA tasks where perplexity serves as a reliable proxy. The fundamental issue is that code generation requires exact syntactic correctness, where semantic coherence (captured by perplexity) does not necessarily translate to functional correctness, limiting perplexity's utility as a selection criterion for code tasks.

\subsubsection{Robust analysis across retrieval methods} \hfill \\

\begin{figure*}[ht]
    \centering
    \includegraphics[width=0.9\textwidth]{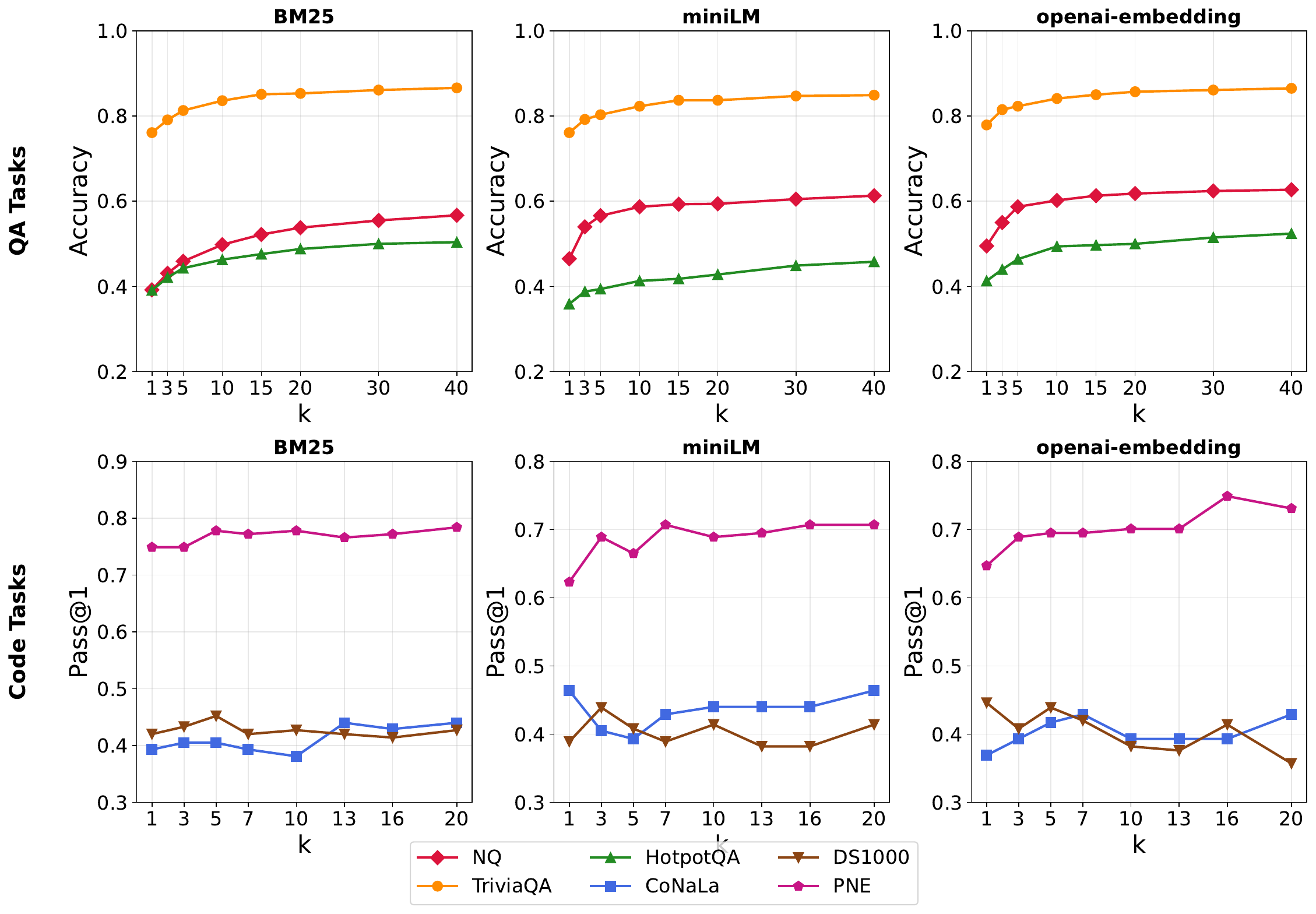}
    \caption{Robustness analysis of RAG performance with varying numbers of documents across different retrieval methods. We adopt two additional retrieval methods, BM25 (sparse) and miniLM (dense), to compare with the openai-embedding baseline. The observed trends are consistent across all methods, demonstrating that our findings regarding optimal $k$ are robust to the choice of retriever.}
    \label{fig:retriever_robustness_docnum}
\end{figure*}

To determine whether our conclusions are artifacts of a specific commercial embedding model or fundamental properties of the tasks themselves, we extended our evaluation to include two distinct retrieval paradigms:
\begin{itemize}
    \item BM25 (Sparse)~\cite{BM25}: A probabilistic information retrieval model that ranks documents based on exact keyword matching and term frequency-inverse document frequency (TF-IDF) saturation. It serves as a rigorous baseline for lexical retrieval.
    \item miniLM (Dense)~\cite{sentence-bert}: A distilled transformer-based model (specifically \textit{all-MiniLM-L6-v2}) that maps queries and documents into a shared dense vector space. It represents the state-of-the-art in efficient, open-source semantic search.
\end{itemize}
We utilized GPT-4o-mini as the fixed generator to isolate the impact of the retrieval method. The comparative results are presented in Figure~\ref{fig:retriever_robustness_docnum}.

For question answering datasets (NQ, TriviaQA, HotpotQA), the performance trajectories of BM25 and miniLM closely mirror those of openai-embedding. Regardless of whether the retrieval signal is lexical (BM25) or semantic (miniLM), accuracy exhibits the characteristic logarithmic growth pattern: a sharp rise within the $k \in [1, 5]$ range, followed by diminishing returns and saturation around $k=10$. This cross-model consistency confirms that the ``sweet spot'' of $k=5-10$ is a fundamental property of the RAG inference process for QA tasks, independent of the underlying retrieval mechanism.
Crucially, the trends for code generation tasks (CoNaLa, DS1000, PNE) further corroborate our earlier findings. As shown in the bottom row of Figure~\ref{fig:retriever_robustness_docnum}, neither BM25 nor miniLM yields a consistent performance gain as context expands. The performance curves remain largely flat or erratic, matching the behavior of the commercial baseline. This demonstrates that the insensitivity to expanded context in code tasks is a robust phenomenon rather than a specific failure of the embedding model.

We posit that this robustness stems from the structural similarity of the retrieval outputs. Regardless of the architecture used (sparse vs. dense), the retriever effectively functions as a filter that returns a ranked list containing a mixture of golden documents and distractors. The downstream performance is primarily driven by the ratio of relevant information to noise within this list. Since the retrieval recall levels for similar tasks remain comparable across standard retrievers—showing rapid saturation for QA but strictly limited recall for code—the resulting trade-off between information gain and noise remains unchanged. Consequently, the conclusions regarding $k$ optimization are retriever-independent.

\begin{tcolorbox}
    \textbf{Answer to RQ2}: 
    Consistent optimal document selection exists for QA tasks with consistent patterns across datasets and models, while code generation generally exhibits statistical insensitivity to expanded context driven by insufficient retrieval recall. For QA tasks, practitioners can reliably use document number between 5 and 10 as a starting point and leverage perplexity as a deployment-friendly proxy for optimization without labeled data. Code generation, however, requires validating retrieval recall sufficiency, as expanded context proves statistically ineffective in low-recall regimes. This fundamental difference suggests that RAG system design should be task-aware, with QA benefiting from standardized approaches while code generation demands more sophisticated, adaptive strategies.
\end{tcolorbox}

\section{RQ3: How should retrieved knowledge be integrated?}
\label{sec:RQ3}

To address this critical engineering decision, we conduct a systematic empirical comparison of representative prompting methods across different categories and resource requirements. This investigation provides practitioners with evidence-based guidance for selecting appropriate prompting strategies based on their specific performance needs and computational constraints.

\subsection{Experimental Design}

\subsubsection{Systematic prompting methods collection and implementation} \hfill \\ 

We categorize existing prompting approaches into four main types through comprehensive literature analysis: \textit{Prompt Tuning}, \textit{Thought Generation}, \textit{Decomposition}, and \textit{Content Verification}~\cite{PromptPapers, PromptReport, chang2024efficientSurvey, vatsal2024survey}.
From each category, we select two representative methods—one few-shot and one zero-shot approach—based on their suitability for RAG scenarios and citation frequency in the literature~\cite{PromptPapers, PromptReport}.
This selection strategy ensures coverage of both resource-intensive approaches (few-shot methods requiring example demonstrations) and resource-efficient approaches (zero-shot methods with minimal computational overhead), addressing the practical trade-off between performance and implementation costs that practitioners face.

Our collection and implementation process follows these steps:

\begin{enumerate}
    \renewcommand{\itemsep}{0.5mm}
    \item \textbf{Literature Review:} We systematically review existing prompting surveys~\cite{PromptReport, sahoo2024systematic, PromptPapers} to identify applicable techniques for RAG contexts.
    \item \textbf{Categorization:} We classify these collected techniques into four categories adapted from established taxonomies~\cite{PromptReport}, ensuring comprehensive coverage of major prompting paradigms.
    \item \textbf{Selection:} We select the most suitable and highly cited few-shot and zero-shot methods from each category, prioritizing techniques with demonstrated effectiveness in similar contexts.
    \item \textbf{Implementation:} We manually implement each prompting method following the original papers' specifications, with three experienced researchers (each with 3+ years ML experience and 1+ years LLM experience) independently reviewing each implementation for fidelity to the original technique and effectiveness for our specific tasks.
\end{enumerate}

An overview of the four categories and selected methods is presented in Tables~\ref{tab:PromtingCategories} \& \ref{tab:PromtIntroduce}.

\begin{center}
\begin{table*}
    \centering
    \small
    \caption{Summary of prompting techniques categories. 
    }
    \begin{tabular}{lcc}
        \toprule
        \textbf{Category}  & \textbf{Category Description} & \textbf{Selected Methods} \\
        \midrule
        \multirow{3}{*}{\textbf{Prompt Tuning}}  & \multirow{3}{8cm}{Enhance LLM's performance by crafting, searching, or generating a better version of prompt from the original prompt to better demonstrate the task to LLM.} & \multirow{3}{*}{\parbox[m]{3cm}{\centering Few-shot~\cite{few-shot}, \\ Emotion Prompting~\cite{emotionPrompting}}}   \\
                                                \\
                                                 \\

        \midrule
        \multirow{2}{*}{\textbf{Thought Generation}}  & \multirow{2}{8cm}{Enhance LLM's reasoning ability by letting it generate intermediate reasoning thoughts.} & \multirow{2}{*}{\parbox{3cm}{\centering Chain-of-Thought~\cite{CoT}, \\ Zero-shot CoT~\cite{zeroshotCOT}}}  \\
                                                    \\

        \midrule
        \multirow{2}{*}{\textbf{Decomposition}}  & \multirow{2}{8cm}{Improve LLM's problem-solving ability by asking LLM to decompose the problem into sub-problems and solve them}  & \multirow{2}{*}{\parbox{3cm}{\centering Least-to-Most~\cite{least-to-most}, \\ Plan-and-Solve~\cite{plan-and-solve}}} \\
                                               \\

        \midrule
        \multirow{2}{*}{\textbf{Content Verification}}  & \multirow{2}{8cm}{Enhance LLM by letting it check and verify the input or output content.} & \multirow{2}{*}{\parbox{3cm}{\centering Self-Refine~\cite{self-refine}, \\ Chain-of-Note~\cite{chain-of-note}}}  \\
                                    \\

        \bottomrule
    \end{tabular}
    
    \label{tab:PromtingCategories}
\end{table*}
\end{center}

\begin{center}
\begin{table}
    \centering
    \small
    \caption{Core idea of selected prompting techniques. 
    }
    \begin{tabular}{ll}
        \toprule
        \textbf{Method} & \textbf{Description} \\
        \midrule
        \multirow{2}{*}{\textbf{Few-shot}}   & \multirow{2}{10cm}{Provide a few exemplars to guide the LLM on approaching similar problems.}  \\
                                    \\

        \midrule
        \multirow{2}{*}{\textbf{Emotion Prompting}}   & \multirow{2}{10cm}{Enhance LLM performance by adding emotional stakes or urgency to prompts.}  \\
                                            \\
        \midrule
        \multirow{2}{*}{\textbf{Chain-of-Thought}}   & \multirow{2}{10cm}{Solve the problem step-by-step, using intermediate reasoning steps.}  \\
                                                \\
                                                    
        \midrule
        \multirow{2}{*}{\textbf{Zero-shot CoT}}   & \multirow{2}{10cm}{Let LLM to solve the problem step-by-step, by instruction rather than exemplars.}  \\
                                                  \\

        \midrule
        \multirow{2}{*}{\textbf{Least-to-Most}}   & \multirow{2}{10cm}{Break down the problem into sub-problems, then solve each of them to reach the final solution.}  \\
                                                  \\

        \midrule
        \multirow{2}{*}{\textbf{Plan-and-Solve}}   & \multirow{2}{10cm}{Develop a plan for solving the problem, then execute the plan.}  \\
                                                  \\

        \midrule
        \multirow{2}{*}{\textbf{Self-Refine}}   & \multirow{2}{10cm}{Generate feedback on the initial response and refine it based on the feedback.}  \\
                                                  \\

        \midrule
        \multirow{2}{*}{\textbf{Chain-of-Note}}   & \multirow{2}{10cm}{Take notes on each retrieved document, assess document relevance to the query, then solve the problem based on relevant notes.}  \\
                                                  \\

        \bottomrule
    \end{tabular}
    \vspace{-10pt}
    \label{tab:PromtIntroduce}
\end{table}
\end{center}

\vspace{-30pt}

\subsubsection{Prompting methods evaluation} \hfill \\


All selected methods are evaluated under identical experimental conditions using the same datasets, models, and RAG performance metrics established in section~\ref{sec:exp_setup}. This controlled setup eliminates the confounding variables present in existing literature, where prompting methods are typically assessed in isolation using different evaluation frameworks.
We compare methods across two key dimensions:
\begin{itemize}
\item Performance: RAG system accuracy and effectiveness using standard evaluation metrics
\item Resource Requirements: Implementation complexity and computational overhead (few-shot vs. zero-shot classification)
\end{itemize}

This systematic investigation addresses a critical gap in RAG research. Existing work predominantly focuses on retrieval optimization while leaving the prompting component largely unexplored.
Our controlled comparative approach evaluates representative methods across four major categories under identical conditions, establishing evidence-based guidelines that account for both performance effectiveness and resource constraints. This enables practitioners to make informed engineering decisions based on their specific deployment requirements and contributes foundational empirical evidence for advancing RAG systems from experimental prototypes toward production-ready solutions.

\subsection{Results and Findings}

\definecolor{out}{RGB}{210, 240, 210}  
\definecolor{und}{RGB}{245, 210, 210} 

\begin{table}
\centering
\captionsetup{width=0.9\textwidth}
\small
\setlength{\tabcolsep}{4pt}
\caption{RAG System Performance with Llama2-13B and various types of prompting methods. Methods outperforming the baseline zero-shot prompt are highlighted in green, while under-performing methods are highlighted in red.
}
\begin{tabular}{l|ccc|ccc}
\toprule
\multirow{2}{*}{Prompting Method} & \multicolumn{3}{c}{QA Tasks} & \multicolumn{3}{c}{Code Tasks} 
\\
\cmidrule(lr){2-4} \cmidrule(lr){5-7}
 & \textbf{NQ} & \textbf{TriviaQA} & \textbf{HotpotQA} & \textbf{CoNaLa} & \textbf{DS1000} & \textbf{PNE} \\
\midrule
\textbf{Zero-Shot}           & 0.559 & 0.825 & 0.443   & 0.214                & 0.134   & 0.557           \\
\midrule
\textbf{Few-Shot}            & \cellcolor{und}0.448 & \cellcolor{und}0.764 & \cellcolor{und}0.350   & \cellcolor{out}0.357   & \cellcolor{out}0.229    &  \cellcolor{out}0.689  \\
\textbf{Emotion Prompting}    & \cellcolor{und}0.557  &  \cellcolor{und}0.816 & \cellcolor{und}0.423     & \cellcolor{und}0.190  & \cellcolor{und}0.108  &   \cellcolor{out}0.575 \\
\midrule
\textbf{Chain-of-Thought}    & \cellcolor{und}0.455 & \cellcolor{und}0.760 & \cellcolor{und}0.369  & \cellcolor{out}0.417    & \cellcolor{out}0.185   &    \cellcolor{out}0.563          \\
\textbf{Zero-Shot CoT}    &    \cellcolor{und}0.542      &  \cellcolor{und}0.810     &   \cellcolor{und}0.426       & \cellcolor{und}0.238   & \cellcolor{und}0.076    &   \cellcolor{und}0.431        \\
\midrule
\textbf{Least-to-Most}       & \cellcolor{und}0.483 & \cellcolor{und}0.779 & \cellcolor{und}0.402  & \cellcolor{out}0.405   & \cellcolor{out}0.153    &  \cellcolor{und}0.473 \\
\textbf{Plan-and-Solve}      & \cellcolor{und}0.400 & \cellcolor{und}0.576 &  \cellcolor{und}0.259  & \cellcolor{und}0.214   & \cellcolor{und}0.045    &   \cellcolor{und}0.359       \\
\midrule
\textbf{Self-Refine}         & \cellcolor{und}0.431 & \cellcolor{und}0.735 & \cellcolor{und}0.297  & \cellcolor{out}0.333    & \cellcolor{und}0.096    &   \cellcolor{und}0.479        \\
\textbf{Chain-of-Note}       & \cellcolor{und}0.482 & \cellcolor{und}0.760 & \cellcolor{und}0.303  & \cellcolor{und}0.119    & \cellcolor{und}0.064    &    \cellcolor{und}0.353       \\
\bottomrule
\end{tabular}
\label{tab:prompt_llama_old}
\end{table}

\begin{table}
\centering
\small
\captionsetup{width=0.9\textwidth}
\setlength{\tabcolsep}{4pt}
\caption{RAG System Performance with GPT-3.5-turbo and various types of prompting methods. Methods outperforming the baseline zero-shot prompt are highlighted in green, while under-performing methods are highlighted in red.
}
\begin{tabular}{l|ccc|ccc}
\toprule
\multirow{2}{*}{Prompting Method} & \multicolumn{3}{c}{QA Tasks} & \multicolumn{3}{c}{Code Tasks} \\
\cmidrule(lr){2-4} \cmidrule(lr){5-7}
 & \textbf{NQ} & \textbf{TriviaQA} & \textbf{HotpotQA} & \textbf{CoNaLa} & \textbf{DS1000} & \textbf{PNE} \\
\midrule
\textbf{Zero-Shot}      & 0.543    & 0.818     & 0.423        & 0.333         & 0.261          &  0.659 \\
\midrule
\textbf{Few-Shot}    & \cellcolor{und}0.492   & \cellcolor{und}0.775     & \cellcolor{und}0.406          & \cellcolor{out}0.476   & \cellcolor{und}0.242    & \cellcolor{out}0.754  \\
\textbf{Emotion Prompting}   & \cellcolor{und}0.541 & \cellcolor{und}0.816 & \cellcolor{und}0.419              & \cellcolor{out}0.393     & \cellcolor{out}0.299   & \cellcolor{und}0.641          \\
\midrule
\textbf{Chain-of-Thought}    & \cellcolor{und}0.510  & \cellcolor{und}0.797  & \cellcolor{out}0.434         & \cellcolor{out}0.440     & \cellcolor{und}0.217   & \cellcolor{out}0.701 \\
\textbf{Zero-Shot CoT}       &  \cellcolor{und}0.531  &  \cellcolor{und}0.812  & \cellcolor{und}0.416            & \cellcolor{out}0.345        & \cellcolor{out}0.280   & \cellcolor{und}0.629  \\
\midrule
\textbf{Least-to-Most}       & \cellcolor{und}0.528    & \cellcolor{und}0.818  & \cellcolor{out}0.442       & \cellcolor{out}0.381          & \cellcolor{und}0.217  & \cellcolor{und}0.659  \\
\textbf{Plan-and-Solve}      & \cellcolor{und}0.495    & \cellcolor{und}0.749  & \cellcolor{out}0.431      & \cellcolor{out}0.405         & \cellcolor{und}0.255  & \cellcolor{und}0.647  \\
\midrule
\textbf{Self-Refine}         & \cellcolor{und}0.530    & \cellcolor{und}0.809   & \cellcolor{und}0.406         & \cellcolor{out}0.429        & \cellcolor{und}0.248  & \cellcolor{und}0.665 \\
\textbf{Chain-of-Note}       & \cellcolor{out}0.553    & \cellcolor{und}0.788   & \cellcolor{und}0.416         & \cellcolor{out}0.405          & \cellcolor{und}0.191 & \cellcolor{und}0.581   \\
\bottomrule
\end{tabular}
\label{tab:prompt_gpt_old}
\end{table}

\begin{table}

\centering
\small
\setlength{\tabcolsep}{4pt}
\captionsetup{width=0.9\textwidth}
\caption{RAG system performance with GPT-4o-mini and various types of prompting methods. Methods outperforming the baseline zero-shot prompt are highlighted in green, while under-performing methods are highlighted in red.
}
\begin{tabular}{l|ccc|ccc}
\toprule
\multirow{2}{*}{Prompting Method} & \multicolumn{3}{c}{QA Tasks} & \multicolumn{3}{c}{Code Tasks} \\
\cmidrule(lr){2-4} \cmidrule(lr){5-7}
 & \textbf{NQ} & \textbf{TriviaQA} & \textbf{HotpotQA} & \textbf{CoNaLa} & \textbf{DS1000} & \textbf{PNE} \\
\midrule
\textbf{Zero-Shot}           & 0.602    & 0.841     & 0.494      & 0.417          & 0.439          & 0.695 \\
\midrule
\textbf{Few-Shot}            & \cellcolor{und}0.58    & \cellcolor{out}0.860     & \cellcolor{und}0.448      & \cellcolor{out}0.500   & \cellcolor{und} 0.229 & \cellcolor{out}0.731 \\
\textbf{Emotion Prompting}   & \cellcolor{out}0.611           & \cellcolor{out}0.843        &  \cellcolor{out}0.506          & \cellcolor{out}0.464   & \cellcolor{und} 0.376  &  \cellcolor{und}0.695        \\
\midrule
\textbf{Chain-of-Thought}    & \cellcolor{und}0.563    & \cellcolor{und}0.837     & \cellcolor{und}0.463      & \cellcolor{out}0.464   & \cellcolor{und} 0.185  & \cellcolor{out}0.766   \\
\textbf{Zero-Shot CoT}      & \cellcolor{out}0.611     & \cellcolor{und}0.841        & \cellcolor{und}0.487            & \cellcolor{out}0.452   & \cellcolor{und} 0.369   &  \cellcolor{und}0.659 \\
\midrule
\textbf{Least-to-Most}       & \cellcolor{und}0.553     & \cellcolor{out}0.844    & \cellcolor{und}0.481      & \cellcolor{und}0.405    & \cellcolor{und}0.204  & \cellcolor{out}0.760 \\
\textbf{Plan-and-Solve}      & \cellcolor{und}0.571      & \cellcolor{out}0.843   & \cellcolor{out}0.503      & \cellcolor{out}0.452    & \cellcolor{und}0.376   & \cellcolor{und}0.659  \\
\midrule
\textbf{Self-Refine}         & \cellcolor{out}0.622     & \cellcolor{out}0.856   & \cellcolor{und}0.491   & \cellcolor{und}0.357   & \cellcolor{und}0.159   &  \cellcolor{out}0.701 \\
\textbf{Chain-of-Note}       & \cellcolor{und}0.561       & \cellcolor{out}0.854    & \cellcolor{out}0.495      & \cellcolor{und}0.310   & \cellcolor{und}0.261   & \cellcolor{und}0.653     \\
\bottomrule
\end{tabular}
\label{tab:prompt_gpt_new}
\end{table}

Table~\ref{tab:prompt_llama_old} \& \ref{tab:prompt_gpt_old} \& \ref{tab:prompt_gpt_new} present RAG system performance with different prompting methods across models and datasets. We use zero-shot instruction prompting as baseline, with outperforming methods highlighted in green and under-performing methods in red. The data reveal that prompting strategies yield highly divergent impacts depending on the model and dataset. For instance, the RAG system with Llama2-13B shows no benefit from any prompting strategy on the QA dataset, whereas all prompting methods improve its performance on the CoNaLa dataset. We analyze these specific variations in detail below.

\subsubsection{Finding-1: prompting methods show limited benefits for QA tasks but substantial improvements for code tasks} \hfill \\

\textbf{Prompting methods demonstrate contrasting effectiveness between QA and code tasks.}
Our evaluation reveals clear performance difference across task types. For QA tasks, prompting methods provide minimal benefits: no methods improve RAG performance with Llama2-13B (Table~\ref{tab:prompt_llama_old}), and only 4 out of 24 configurations show improvement with GPT-3.5-turbo (Table~\ref{tab:prompt_gpt_old}). Even with the more capable GPT-4o-mini, improvements remain marginal—the best result achieves only 2\% accuracy improvement using self-refine on the NQ dataset, while many enhancements are negligible (e.g., Chain-of-Note improves HotpotQA accuracy by merely 0.1\%), as shown in Table~\ref{tab:prompt_gpt_new}.
In contrast, code generation tasks demonstrate substantial improvements from prompting methods. Chain-of-Thought increases pass@1 by over 0.2 on CoNaLa with Llama2-13B, while Few-shot prompting improves pass@1 by approximately 0.1 on PNE with GPT-3.5-turbo. These gains represent meaningful performance enhancements that justify the additional prompting overhead.

This performance disparity stems from fundamental task characteristics. QA tasks—including multi-hop variants—primarily involve straightforward factual retrieval with limited reasoning requirements, rendering Thought Generation methods (e.g., Chain-of-Thought) and Decomposition approaches (e.g., Plan-and-Solve) largely ineffective. Furthermore, QA tasks present explicit questions with clear information needs, eliminating the problem comprehension challenges that Prompting Tuning methods are designed to address.
Code generation tasks, however, require synthesizing retrieved API information into executable code (as established in RQ1), creating a natural fit for structured prompting approaches that guide reasoning and planning processes. The complexity of translating natural language requirements into functional code benefits from the intermediate reasoning steps that prompting methods provide.

These findings suggest that effective RAG system optimization should adopt task-specific strategies: prioritizing retrieval quality improvements for QA tasks while emphasizing reasoning-enhanced prompting techniques for code generation tasks.

For the remaining analysis, we examine both QA and code generation tasks but emphasize code generation due to the overall lack of effect of prompting methods for QA tasks.

\subsubsection{Finding-2: model capability determines prompting strategy effectiveness} \hfill \\

\textbf{For code generation tasks, Few-shot methods significantly benefit weaker models, while zero-shot methods have negative effects.}
For Llama2-13B, few-shot prompting methods consistently improve RAG system performance, with improvements up to 85\% for CoNaLa using Chain-of-Thought prompting.
Conversely, zero-shot methods consistently reduce code generation accuracy for Llama2-13B, as shown by the red-colored results in Table~\ref{tab:prompt_llama_old}. The performance gap with few-shot and zero-shot methods in the same category is dramatic, commonly around 2x performance gap, with up to 340\% for DS1000 using Least-to-Most (few-shot-style method) and Plan-and-Solve (zero-shot style method).
This suggests that instruction-only prompting without examples can overwhelm weaker models in the RAG context.

\medskip
\textbf{Advanced models have convergent performance between strategies.}
Unlike the pattern observed with Llama2-13B, advanced models (GPT-3.5-turbo and GPT-4o-mini) show minimal performance gaps between few-shot and zero-shot approaches. 
Especially for GPT-4o-mini, few-shot and zero-shot methods in the same category perform quite similarly on CoNaLa, and zero-shot methods even provide more benefits in CoNaLa Least-to-Most and all DS1000 scenarios.
Remarkably, simple instruction additions (e.g., ``this is very important to my career'' for emotion prompting) can substantially improve performance, boosting CoNaLa performance from 41.7\% to 46.4\% in pass rate.

For QA tasks, the same pattern holds: even though prompting methods generally do not provide benefits, RAG with Llama2-13B shows performance degradation when incorporating prompting methods, while RAG with more advanced GPT-4o-mini shows some improved cases, though limited.

\medskip
\textbf{Weaker models can outperform advanced models under specific prompting conditions for code tasks.}
In certain cases, when given identical prompts, RAG systems with weaker models can surpass those with advanced models.
For instance, with the exact same Least-to-Most prompting, the RAG system with Llama2-13B performs better than GPT-3.5-turbo and achieves similar performance to GPT-4o-mini on the CoNaLa dataset.
This phenomenon emphasizes that prompting methods must be considered in RAG scenarios, and that model capability and prompting strategy interactions need to be evaluated together rather than independently.

\medskip
\textbf{Optimal prompting can enable weaker models to outperform advanced models with suboptimal prompts.}
A particularly striking finding emerges when comparing optimally-prompted weaker models against advanced models with base or suboptimal prompting. For instance, Llama2-13B with Chain-of-Thought prompting outperforms GPT-3.5-turbo with base prompting and achieves comparable performance to GPT-4o-mini on CoNaLa dataset, and similar results occur for Llama2-13B with Few-shot prompting on PNE dataset. For instance, with few-shot prompting, the RAG system with Llama2-13B generates valid code for a problem in the PNE dataset. However, with Zero-shot CoT, the system with GPT-4o-mini incorrectly uses `find\_repeats' from the API documents and generates an invalid code snippet, as shown in Figure~\ref{fig:LlamaOutperformGPT}.
This suggests that prompting strategy can be more influential than raw model capability in RAG contexts for code generation tasks. The implication is significant: carefully designed prompting strategies may offer a cost-effective alternative to deploying more expensive, advanced models, particularly when prompt optimization receives adequate attention.

\begin{figure}
\centering
\begin{tikzpicture}[scale=0.85, transform shape]

\node[draw, rounded corners=10pt, fill=questionbg, text width=15cm, align=left] (code_question) at (0.2,0) {
\small \textbf{Problem Description:} Find the most frequent number in a NumPy array and return the number
};

\node[draw, rounded corners=10pt, fill=answerbg, text width=15cm, align=left] (code_answer1) at (0.2,-3.1) {
\small \textbf{Correct Generated Code when using Llama2-13B and Few-shot Prompting:}
\small \begin{lstlisting}[language=Python, keywordstyle={}, numbers=none, emph={}, emphstyle=\color{red}, basicstyle=\ttfamily]
## Example1 ...
## Example2 ...
## Example3 ...

## API Documents: ...
## Incomplete Code: ...
## Complete Code:
import numpy as np
def find_most_frequent_number(arr):
    # Find the most frequent number in a NumPy array
    # Return the number
    unique, counts = np.unique(arr, return_counts=True)
    return unique[np.argmax(counts)]
\end{lstlisting}
};

\node[draw, rounded corners=10pt, fill=wronganswerbg, text width=15cm, align=left] (code_answer2) at (0.2,-9.9) {
\small \textbf{Incorrect Generated Code when using GPT-4o-mini and Zero-shot CoT Prompting:}
\small \begin{lstlisting}[language=Python, keywordstyle={}, numbers=none, emph={find_repeats}, emphstyle=\color{red}, basicstyle=\ttfamily]
## API Documents: ...
## Incomplete Code: ...

Lets think it step by step.
To complete the `find_most_frequent_number` function, we can utilize the `find_repeats`
function from the API documents. This function will help us identify the repeated values
and their counts. We will then return the value with the highest count.
Here is the completed code:

<code>
import numpy as np
def find_most_frequent_number(arr):
    # Find the most frequent number in a NumPy array
    repeats, counts = np.stats.find_repeats(arr)
    if counts.size == 0:
        return None  # No repeats found
    most_frequent_index = np.argmax(counts)
    return repeats[most_frequent_index]
</code>
\end{lstlisting}
};


\end{tikzpicture}
\caption{A specific case demonstrating that optimal prompting can enable weaker models to outperform advanced models with suboptimal prompts. Although Llama2-13B significantly underperforms compared to GPT-4o-mini in general, it produces correct code with proper prompting (Few-shot), whereas GPT-4o-mini with suboptimal prompting (Zero-Shot CoT) fails.}
\label{fig:LlamaOutperformGPT}
\vspace{-10pt}
\end{figure}

\subsubsection{Finding-3: prompting method effectiveness varies dramatically across code tasks.} \hfill \\

\textbf{Prompting methods show most effectiveness in CoNaLa, and minor for DS1000.}
By comparing the effectiveness across datasets, we observe that most prompting methods can improve performance on CoNaLa, while few methods can improve RAG system performance on DS1000. PNE falls in the middle, showing improvement in approximately half of the scenarios.
This matches the pattern of complexity in code dataset, that the DS1000 problems have the highest average API usage, while PNE and CoNaLa has fewer, as shown in Table~\ref{tab:NumOfOracleDocuments}. Additionally, the PNE problems have higher average prompt length than CoNaLa. We illustrate a negative instance of this in Figure~\ref{fig:CoTFailure}, where advanced prompting proves detrimental. Specifically, while the base RAG system generates correct code, applying Chain-of-Thought prompting causes the LLM to ``overthink'' and implement an unrequested clustering step, resulting in a failure.
This pattern suggests that task complexity creates natural boundaries for prompting effectiveness.

\begin{table}[t]
    \centering
    \caption{Complexity comparison across code generation datasets measured by solution size, API document numbers, and prompt length. 
    }
    \renewcommand{\tabcolsep}{7pt}
    \small
    \begin{tabular}{lccc}
        \toprule
        \textbf{Metric} & \textbf{CoNaLa} & \textbf{DS1000} & \textbf{PNE}\\
        \midrule
        \textbf{Avg. Code Lines in Solutions}      & 1.0 & \textbf{5.5} & 1.17 \\
        \textbf{Avg. API Documents}   & 1.74 & \textbf{2.68} & 1.49 \\
        \textbf{Avg. Length of Zero-shot Prompt}   & 484.8 & \textbf{3139.5}  & 1805.8 \\
        \bottomrule
    \end{tabular}
    \vspace{-10pt}
    \label{tab:NumOfOracleDocuments}
\end{table}

\subsubsection{Finding-4: prompting method categories have different effectiveness for code tasks.} \hfill \\

\begin{table}
\centering
\small
\setlength{\tabcolsep}{3pt}
\caption{Prediction distribution analysis: percentage of samples correct only in each method (baseline only / advanced only).
}
\begin{tabular}{l|l|ccc|ccc}
\toprule
\multirow{2}{*}{Model} & \multirow{2}{*}{Prompting Method} & \multicolumn{3}{c}{QA Tasks} & \multicolumn{3}{c}{Code Tasks} \\
\cmidrule(lr){3-5} \cmidrule(lr){6-8}
 & & \textbf{NQ} & \textbf{TriviaQA} & \textbf{HotpotQA} & \textbf{CoNaLa} & \textbf{DS1000} & \textbf{PNE} \\
\midrule

\multirow{8}{*}{\textbf{Llama2-13B}} 
& \textbf{Few-Shot}          & 15.1\% / 4.0\%*    & 8.2\% / 2.1\%*     & 13.0\% / 3.7\%*     & 8.3\% / 22.6\%*        & 3.8\% / 13.4\%*        & 7.8\% / 21.0\%* \\
& \textbf{Emotion Prompting} & 2.0\% / 1.8\%*    & 1.5\% / 0.5\%*  & 3.3\% / 1.3\%*      & 2.4\% / 0.0\%         & 5.1\% / 2.5\%         & 2.4\% / 4.2\% \\

& \textbf{Chain-of-Thought}  & 14.4\% / 4.0\%*    & 8.1\% / 1.6\%*   & 13.4\% / 6.0\%*      & 6.0\% / 26.2\%*        & 6.4\% / 11.5\%*       & 15.6\% / 16.2\%* \\
& \textbf{Zero-Shot CoT}     & 5.8\% / 4.1\%*    & 3.4\% / 1.9\%*  & 4.9\% / 3.2\%      & 1.2\% / 3.6\%         & 9.6\% / 3.8\%*         & 19.8\% / 7.2\%* \\

& \textbf{Least-to-Most}   & 21.5\% / 5.7\%*    & 7.1\% / 2.5\%*     & 11.2\% / 7.0\%*      & 11.9\% / 31.0\%*       & 7.0\% / 8.9\%*         & 21.6\% / 13.2\%* \\
& \textbf{Plan-and-Solve}    & 17.2\% / 4.5\%*    & 28.4\% / 3.5\%   & 23.4\% / 5.1\%*     & 11.9\% / 11.9\%*       & 10.8\% / 1.9\%        & 25.7\% / 6.0\%* \\

& \textbf{Self-Refine}     & 17.2\% / 4.5\%*    & 11.8\% / 2.9\%*     & 19.5\% / 5.0\%*      & 3.6\% / 15.5\%        & 5.7\% / 1.9\%         & 14.4\% / 6.6\%* \\
& \textbf{Chain-of-Note}   & 12.5\% / 4.9\%*    & 9.6\% / 3.1\%*     & 19.1\% / 5.2\%*      & 14.3\% / 4.8\%        & 8.3\% / 1.3\%         & 24.6\% / 4.2\%* \\
\midrule

\multirow{8}{*}{\textbf{GPT-3.5-turbo}} 
& \textbf{Few-Shot}       & 10.4\% / 5.4\%*    & 7.7\% / 3.5\%*     & 7.1\% / 5.4\%*      & 10.7\% / 25.0\%*      & 12.7\% / 10.8\%*        & 4.2\% / 13.8\%* \\
& \textbf{Emotion Prompting}   & 2.6\% / 2.5\%*   & 1.5\% / 1.3\%*     & 2.9\% / 2.6\%*      & 4.8\% / 10.7\%         & 3.8\% / 7.6\%*          & 4.2\% / 2.4\% \\

& \textbf{Chain-of-Thought} & 8.6\% / 5.3\%*    & 6.0\% / 4.1\%*     & 7.6\% / 8.7\%*      & 10.7\% / 21.4\%*      & 12.1\% / 7.6\%*         & 8.4\% / 12.6\%* \\
& \textbf{Zero-Shot CoT}   & 2.6\% / 1.5\%*    & 2.0\% / 1.5\%*     & 3.4\% / 2.7\%*      & 4.8\% / 6.0\%          & 4.5\% / 6.4\%*          & 7.2\% / 4.2\%* \\

& \textbf{Least-to-Most}   & 8.5\% / 7.0\%*   & 5.0\% / 5.0\%*     & 8.2\% / 10.1\%*      & 10.7\% / 15.5\%*        & 12.7\% /8.3\%*         & 13.2\% / 12.0\%* \\
& \textbf{Plan-and-Solve}   & 11.3\% / 6.5\%*    & 11.0\% / 4.2\%*     & 8.4\% / 9.2\%*     & 3.6\% / 10.7\%         & 7.0\% / 6.4\%*          & 6.0\% / 4.8\%* \\

& \textbf{Self-Refine}     & 8.2\% / 6.9\%*    & 5.9\% / 5.1\%*     & 8.6\% / 6.9\%*      & 10.7\% / 20.2\%*        & 8.9\% / 7.6\%*          & 13.2\% / 13.8\%* \\
& \textbf{Chain-of-Note}    & 6.7\% / 7.7\%*    & 7.1\% / 4.3\%*     & 8.7\% / 8.1\%*     & 7.1\% / 14.3\%*         & 13.4\% /6.4\%*         & 16.8\% / 9.0\%* \\
\midrule

\multirow{8}{*}{\textbf{GPT-4o-mini}} 
& \textbf{Few-Shot}     & 5.7\% / 3.4\%*    & 1.5\% / 3.4\%*     & 8.0\% / 3.3\%*   & 7.1\% / 15.5\%*   & 29.9\% / 8.9\%*      & 3.0\% / 6.6\%* \\
& \textbf{Emotion Prompting}   & 1.3\% / 2.1\%*    & 0.8\% / 0.9\%*     & 1.2\% / 2.4\%*   & 0\% / 4.8\%      & 9.6\% / 3.2\%*       & 1.8\% / 1.8\% \\

& \textbf{Chain-of-Thought}  & 7.3\% / 3.5\%*    & 2.8\% / 2.4\%*     & 9.0\% / 5.9\%*    & 7.1\% / 11.9\%*   & 33.8\% / 8.3\%*   & 3.0\% /10.2\%* \\
& \textbf{Zero-Shot CoT}   & 1.5\% / 2.3\%*    & 1.1\% / 1.1\%*     & 6.5\% / 5.8\%*     & 1.2\% / 4.8\%    & 11.5\% / 4.5\%*   & 7.2\% /3.6\%* \\

& \textbf{Least-to-Most}   & 9.0\% / 4.2\%*    & 3.0\% / 3.3\%*     & 8.5\% / 7.2\%*      & 7.1\% / 6.0\%*    & 31.8\% / 8.3\%*   & 3.6\% /10.2\%* \\
& \textbf{Plan-and-Solve}  & 7.9\% / 4.8\%*   & 3.9\% / 4.1\%*     & 7.7\% / 8.6\%*      & 2.4\% / 6.0\%       & 13.4\% / 7.0\%*  & 7.8\%/4.2\%* \\

& \textbf{Self-Refine}   & 2.5\% / 4.5\%*    & 1.5\% / 3.0\%*     & 4.3\% / 4.1\%*      & 14.3\% / 8.3\%*   & 33.8\% / 5.7\%*  & 9.0\%/ 9.6\%* \\
& \textbf{Chain-of-Note}  & 8.0\% / 3.9\%*    & 2.4\% / 3.6\%*     & 6.9\% / 7.0\%*      & 14.3\% / 3.6\%     & 24.8\% / 7.0\%*  & 9.6\% /5.4\%* \\

\bottomrule
\end{tabular}
\label{tab:prompt_prediction_distribution}
\vspace{-10pt}
\end{table}

\textbf{Decomposition and Content verification methods provide minor improvement for code tasks.}
While the \textit{Prompt Tuning} and \textit{Thought Generation} categories of prompting methods show at least one beneficial method each in every dataset and model combination, \textit{Decomposition} methods (Least-to-Most and Plan-and-Solve) and \textit{Content Verification} methods (Self-Refine and Chain-of-Note) seldom improve code generation accuracy, they only have positive effect In 6 out of 18, and 4 out of 18 scenarios, while for \textit{Prompt Tuning} and \textit{Thought Generation}, the number is 11 out of 18 and 10 out of 18.
Moreover, regardless of whether they provide benefits or not, the performance of RAG system with those methods often achieves the lowest performance across different models and tasks, especially for \textit{Content Verification} methods. 
For instance, for CoNaLa, DS1000 with Llama2-13B, Chain-of-Note exhibits the lowest performance, while Self-Refine exhibits the lowest performance across few-shot based methods; for DS1000, PNE with GPT-3.5-turbo, Chain-of-Note exhibits the lowest performance; for GPT-4o-mini, Chain-of-Note exhibits the lowest performance in all three datasets, while Self-Refine exhibits the lowest performance across all few-shot methods in three code tasks.
On one hand, this may be due to the higher complexity of such methods; on the other hand, this may be due to the inherent incompatibility of the rationale behind those methods and the code generation tasks, which has also been mentioned in other scenarios~\cite{tony2024promptingSecure}.

\begin{figure}
\centering
\begin{tikzpicture}[scale=0.85, transform shape]

\node[draw, rounded corners=10pt, fill=questionbg, text width=15cm, align=left] (code_question) at (0.2,0) {
\small \textbf{Problem Description:} After clustering a distance matrix with scipy.cluster.hierarchy.linkage, and assigning each sample to a cluster using scipy.cluster.hierarchy.cut\_tree, I would like to extract one element out of each cluster, which is the closest to that cluster's centroid.
};

\node[draw, rounded corners=10pt, fill=answerbg, text width=15cm, align=left] (code_answer1) at (0.2,-3.1) {
\small \textbf{Correct Generated Code when not applying advanced prompting strategy:}
\small \begin{lstlisting}[language=Python, keywordstyle={}, numbers=none, emph={}, emphstyle=\color{red}, basicstyle=\ttfamily]
import numpy as np
import scipy.spatial
centroids = np.random.rand(5, 3)
data = np.random.rand(100, 3)
# Calculate the distances from each data point to each centroid
distances = scipy.spatial.distance.cdist(data, centroids)
# Find the index of the closest point to each centroid
closest_indices = np.argmin(distances, axis=0)
# Extract the closest points using the indices
result = data[closest_indices]
print(result)
\end{lstlisting}
};

\node[draw, rounded corners=10pt, fill=wronganswerbg, text width=15cm, align=left] (code_answer2) at (0.2,-10.6) {
\small \textbf{Incorrect Generated Code when applying Chain of Thought:}
\small \begin{lstlisting}[language=Python, keywordstyle={}, numbers=none, emph={fcluster}, emphstyle=\color{red}, basicstyle=\ttfamily]
To extract the closest element to each clusters centroid after performing clustering, you
can follow these steps:
1. Calculate the centroids of each cluster.
2. For each centroid, find the data point that is closest to it.
Heres how you can implement this in Python using NumPy and SciPy:

import numpy as np
from scipy.cluster.hierarchy import linkage, fcluster
from scipy.spatial.distance import cdist
# Sample data
data = np.random.rand(100, 3)
# Perform hierarchical clustering
Z = linkage(data, method='ward')
clusters = fcluster(Z, t=5, criterion='maxclust')
# Calculate centroids for each cluster
centroids = np.array([data[clusters == i].mean(axis=0)
                      for i in range(1, np.max(clusters) + 1)])
# Find the closest point to each centroid
closest_points = []
for centroid in centroids:
    distances = np.linalg.norm(data - centroid, axis=1) # Calculate distances
    closest_point_index = np.argmin(distances)  # Get the index of the closest point
    closest_points.append(data[closest_point_index])  # Append the closest point
result = np.array(closest_points)
print(result)
\end{lstlisting}
};


\end{tikzpicture}
\caption{A specific case demonstrating that advanced prompting strategies may adversely affect the RAG system. In this instance, the RAG system generates correct code without specific prompting strategies; however, when Chain-of-Thought prompting is applied, the LLM ``overthinks'' and performs an additional clustering operation not requested in the problem statement.}
\label{fig:CoTFailure}
\vspace{-10pt}
\end{figure}

\subsubsection{Finding-5: prompting methods create orthogonal problem-solving patterns for both QA and code.} \hfill \\

\textbf{Advanced prompting methods fundamentally alter which problems get solved rather than simply improving overall accuracy.}
From our results, we observe that a substantial number of prompting methods cannot provide benefits to the RAG system (shown as red-colored data in our tables).
While the prompting methods show significant improvements in other scenarios, they often fail in our experimental setup.
We hypothesize that this occurs for similar reasons as RAG with 100\% golden documents sometimes failing: the extra complexity introduced by prompting methods can cause the RAG system to improperly understand documents, output incorrectly formatted code, use undefined variables, or even forget the task entirely.

To investigate this phenomenon, we conduct prediction distribution analysis to examine samples that are only correct with zero-shot baseline prompting versus those only correct with advanced prompting methods. In Table~\ref{tab:prompt_prediction_distribution}, the first number in each cell shows the percentage of samples only correct with RAG baseline prompts, while the second number shows samples only correct with advanced prompts. Chi-square tests identify significantly different prediction distributions (marked with *).


\medskip
\textbf{Orthogonal effectiveness patterns emerge across all scenarios, particularly pronounced for few-shot style methods.}
Even prompting methods that provide overall benefits cannot solve problems that zero-shot baseline methods can handle. This phenomenon appears everywhere: only for CoNaLa with Llama2-13B, RAG system with Chain-of-Thought prompting additionally solves 26.2\% more samples, but simultaneously fails on a substantial 7.1\% of originally correct samples, while RAG system with Least-to-Most, Self-Refine and Few-Shot has the same patterns.
This orthogonal pattern holds even for methods causing overall performance degradation - they solve different problems rather than simply performing worse.
For example, CoNaLa Plan-and-Solve with Llama2-13B correctly solves 11.9\% samples that are originally wrong, while making mistakes on 13.1\% ones that are originally correct. Similarly, PNE Least-to-Most with GPT-3.5-turbo shows 12.0\% and 13.2\% respectively.

The same orthogonal patterns hold for QA scenarios. Although prompting methods cannot provide benefits or provide only limited benefits for RAG system-based question answering, prediction distributions are actually quite distinct from the baseline prompt. For example, GPT-4o-mini Plan-and-Solve can solve extra 8.6\% cases but fails on 7.7\% cases, and for the less complicated dataset TriviaQA where the LLM has high accuracy, we can also observe that Self-Refine solves 5.1\% more cases, and with Least-to-Most the RAG system can additionally solve 5.0\% more cases. Rather than providing universal improvements, prompting methods fundamentally change which problems get solved, creating opportunities for ensemble approaches that leverage the orthogonal strengths of different methods.

\vspace{-10pt}

\subsubsection{Robust analysis across retrieval methods} \hfill \\

To verify whether the effectiveness of prompting strategies is contingent on the retrieval source, we replicated the RQ3 analysis using BM25 and miniLM alongside the original text-embedding-3-small baseline. The results, detailed in Table~\ref{tab:prompt_robustness_full}, demonstrate that the impact of prompting strategies is largely retriever-independent, with performance patterns remaining consistent across diverse retrieval paradigms.

In the domain of code generation, the interaction between prompting strategies and task difficulty exhibits a uniformity across all three retrievers. For the challenging DS1000 dataset, we observe a consistent ``failure to improve''; regardless of whether the context is retrieved via sparse (BM25) or dense (miniLM/OpenAI) methods, almost all prompting strategies fail to outperform the Zero-Shot baseline. This corroborates our earlier finding that difficult code tasks with strict unit tests are resistant to prompt engineering when the underlying model lacks sufficient reasoning capability. Conversely, for CoNaLa, the trend is universally positive, with most prompting strategies yielding improvements across all retrieval scenarios. Furthermore, specific failure modes in PNE are preserved, where strategies that underperformed with the text-embedding-3-small—specifically \textit{Plan-and-Solve} and \textit{Zero-Shot CoT}—also struggle or degrade performance in the BM25 and miniLM settings. These results confirm that the effectiveness of a prompting strategy is primarily determined by the alignment between task difficulty and model capability (Finding-2), rather than the specific nature of the retrieved context.

A similar stability characterizes the QA tasks, reinforcing the observation that prompting often yields limited benefits for RAG-based QA (Finding-1). On NQ, complex prompting methods such as \textit{Few-Shot}, \textit{Chain-of-Thought}, and \textit{Chain-of-Note} consistently underperform the Zero-Shot baseline across all three retrieval scenarios. While TriviaQA allows for marginal improvements due to high base accuracy, the trends in HotpotQA remain consistent: although the text-embedding-3-small baseline exhibits slightly more frequent performance degradation, the alternative retrievers do not unlock any significant gains from complex prompting.

In conclusion, just as observed in our RQ2 analysis, the interaction between RAG components shows high stability. While the absolute performance values fluctuate based on the retriever's quality, the relative effectiveness of prompting strategies remains identical. A strategy that fails to correct reasoning errors with OpenAI embeddings is also unlikely to succeed with BM25, confirming that our guidelines regarding prompting strategy selection are robust and generalizable across different retrieval architectures.

\begin{table*}[t]
\centering
\small
\setlength{\tabcolsep}{1.2pt} 
\captionsetup{width=0.95\textwidth}

\caption{Robustness of prompting strategies across different retrieval methods. We report performance using BM25 (left), miniLM (middle), and text-embedding-3-small from OpenAI (right) for each dataset. Cells highlighted in \colorbox{out}{green} outperform the Zero-Shot baseline for that specific retriever, while \colorbox{und}{red} indicates underperformance.}
\label{tab:prompt_robustness_full}

\begin{tabular}{l|ccc|ccc|ccc|ccc|ccc|ccc}
\toprule
\multirow{3}{*}{\textbf{Prompting Method}} & \multicolumn{9}{c}{\textbf{QA Tasks}} & \multicolumn{9}{c}{\textbf{Code Tasks}} \\
\cmidrule(lr){2-10} \cmidrule(lr){11-19}
 & \multicolumn{3}{c}{\textbf{NQ}} & \multicolumn{3}{c}{\textbf{TriviaQA}} & \multicolumn{3}{c}{\textbf{HotpotQA}} & \multicolumn{3}{c}{\textbf{CoNaLa}} & \multicolumn{3}{c}{\textbf{DS1000}} & \multicolumn{3}{c}{\textbf{PNE}} \\
 & \small{B} & \small{M} & \small{O} & \small{B} & \small{M} & \small{O} & \small{B} & \small{M} & \small{O} & \small{B} & \small{M} & \small{O} & \small{B} & \small{M} & \small{O} & \small{B} & \small{M} & \small{O} \\
\midrule
\textbf{Zero-Shot} & 0.498 & 0.587 & 0.602 & 0.836 & 0.823 & 0.841 & 0.463 & 0.413 & 0.494 & 0.405 & 0.393 & 0.417 & 0.452 & 0.408 & 0.439 & 0.778 & 0.665 & 0.695 \\
\midrule
\textbf{Few-Shot}          & \cellcolor{und}0.495 & \cellcolor{und}0.573 & \cellcolor{und}0.580 & \cellcolor{out}0.856 & \cellcolor{out}0.845 & \cellcolor{out}0.860 & \cellcolor{und}0.420 & \cellcolor{und}0.377 & \cellcolor{und}0.448 & \cellcolor{out}0.476 & \cellcolor{out}0.500 & \cellcolor{out}0.500 & \cellcolor{und}0.255 & \cellcolor{und}0.236 & \cellcolor{und}0.229 & \cellcolor{out}0.784 & \cellcolor{out}0.749 & \cellcolor{out}0.731 \\

\textbf{Emotion Prompting} & \cellcolor{out}0.503 & \cellcolor{out}0.589 & \cellcolor{out}0.611 & \cellcolor{out}0.836 & \cellcolor{und}0.820 & \cellcolor{out}0.843 & \cellcolor{out}0.477 & \cellcolor{out}0.418 & \cellcolor{out}0.506 & \cellcolor{out}0.512 & \cellcolor{out}0.452 & \cellcolor{out}0.464 & \cellcolor{und}0.439 & \cellcolor{und}0.350 & \cellcolor{und}0.376 & \cellcolor{out}0.778 & \cellcolor{out}0.683 & \cellcolor{out}0.695 \\
\midrule
\textbf{Chain-of-Thought} & \cellcolor{und}0.477 & \cellcolor{und}0.574 & \cellcolor{und}0.563 & \cellcolor{out}0.862 & \cellcolor{out}0.843 & \cellcolor{und}0.837 & \cellcolor{out}0.477 & \cellcolor{out}0.416 & \cellcolor{und}0.463 & \cellcolor{out}0.536 & \cellcolor{out}0.488 & \cellcolor{out}0.464 & \cellcolor{und}0.236 & \cellcolor{und}0.191 & \cellcolor{und}0.185 & \cellcolor{und}0.760 & \cellcolor{out}0.760 & \cellcolor{out}0.766 \\
\textbf{Zero-Shot CoT} & \cellcolor{und}0.496 & \cellcolor{out}0.592 & \cellcolor{out}0.611 & \cellcolor{und}0.834 & \cellcolor{out}0.824 & \cellcolor{out}0.841 & \cellcolor{out}0.479 & \cellcolor{out}0.426 & \cellcolor{und}0.487 & \cellcolor{out}0.488 & \cellcolor{out}0.476 & \cellcolor{out}0.452 & \cellcolor{und}0.446 & \cellcolor{und}0.389 & \cellcolor{und}0.369 & \cellcolor{out}0.784 & \cellcolor{und}0.659 & \cellcolor{und}0.659 \\
\midrule
\textbf{Least-to-Most} & \cellcolor{out}0.518 & \cellcolor{out}0.588 & \cellcolor{und}0.553 & \cellcolor{out}0.867 & \cellcolor{out}0.858 & \cellcolor{out}0.844 & \cellcolor{out}0.486 & \cellcolor{und}0.409 & \cellcolor{und}0.481 & \cellcolor{out}0.488 & \cellcolor{out}0.476 & \cellcolor{und}0.405 & \cellcolor{und}0.242 & \cellcolor{und}0.210 & \cellcolor{und}0.204 & \cellcolor{und}0.760 & \cellcolor{out}0.778 & \cellcolor{out}0.760 \\
\textbf{Plan-and-Solve} & \cellcolor{out}0.500 & \cellcolor{und}0.582 & \cellcolor{und}0.571 & \cellcolor{out}0.889 & \cellcolor{out}0.875 & \cellcolor{out}0.843 & \cellcolor{out}0.516 & \cellcolor{out}0.458 & \cellcolor{out}0.503 & \cellcolor{out}0.488 & \cellcolor{out}0.464 & \cellcolor{out}0.452 & \cellcolor{und}0.420 & \cellcolor{out}0.414 & \cellcolor{und}0.376 & \cellcolor{und}0.731 & \cellcolor{und}0.659 & \cellcolor{und}0.659 \\
\midrule
\textbf{Self-Refine} & \cellcolor{out}0.534 & \cellcolor{out}0.599 & \cellcolor{out}0.622 & \cellcolor{out}0.884 & \cellcolor{out}0.874 & \cellcolor{out}0.856 & \cellcolor{out}0.501 & \cellcolor{out}0.438 & \cellcolor{und}0.491 & \cellcolor{und}0.357 & \cellcolor{out}0.405 & \cellcolor{und}0.357 & \cellcolor{und}0.153 & \cellcolor{und}0.146 & \cellcolor{und}0.159 & \cellcolor{und}0.749 & \cellcolor{out}0.707 & \cellcolor{out}0.701 \\
\textbf{Chain-of-Note} & \cellcolor{und}0.494 & \cellcolor{und}0.575 & \cellcolor{und}0.561 & \cellcolor{out}0.904 & \cellcolor{out}0.888 & \cellcolor{out}0.854 & \cellcolor{out}0.507 & \cellcolor{out}0.443 & \cellcolor{out}0.495 & \cellcolor{out}0.464 & \cellcolor{out}0.417 & \cellcolor{und}0.310 & \cellcolor{und}0.299 & \cellcolor{und}0.280 & \cellcolor{und}0.261 & \cellcolor{und}0.725 & \cellcolor{out}0.689 & \cellcolor{und}0.653 \\
\bottomrule
\end{tabular}
\end{table*}

\begin{tcolorbox}
    \textbf{Answer to RQ3}: RAG systems can be improved through prompting, but success depends critically on task characteristics and model capability alignment. 
    Code generation benefits significantly from prompting methods, while QA tasks show minimal benefits regardless of approach.
    Model capability determines optimal strategies: weaker models require few-shot prompting for code generation but experience degradation on QA tasks, whereas advanced models demonstrate flexibility across strategies.
    Most remarkably, well-designed prompting can enable weaker models to outperform advanced models in code generation.
    Fundamentally, prompting methods create distinct problem-solving pathways rather than simply enhancing existing approaches, highlighting their transformative potential for RAG systems.
\end{tcolorbox}

\section{Discussion}
\label{sec:discussion}

\subsection{Future Directions}

\noindent \textbf{Eliminating RAG system failures.}
Our empirical study reveals several critical causes of RAG system degradation: incorrect utilization of useful documents, inclusion of excessive irrelevant information, and degraded instruction-following capabilities within RAG contexts.
Understanding and mitigating these failure modes is essential for developing high-performance LLM-driven RAG systems. Future work should focus on developing robust monitoring mechanisms and intervention strategies to prevent performance decline in production environments.

\noindent \textbf{Uncertainty evaluation for RAG system.}
For practical RAG-driven applications, uncertainty metrics represent a valuable quality assurance tool. Our study demonstrates a strong correlation between perplexity and RAG system performance on question answering tasks, consistent with prior findings~\cite{flare, huang2023look}.
This correlation highlights perplexity's potential as a performance indicator for monitoring system robustness.
However, the relationship between perplexity and code generation correctness proves less reliable.
While perplexity and other uncertainty-based methods show promise for RAG systems, their effectiveness requires further investigation, particularly for code generation tasks where traditional uncertainty measures may not adequately capture semantic correctness.

\noindent \textbf{Prompting method enhancement.} 
Our findings in RQ3 reveal that appropriate prompting strategies can significantly enhance RAG system performance, while poorly designed prompts can be detrimental.
These results underscore the importance of model-specific, task-specific prompt optimization. Future research should develop systematic frameworks for prompt engineering that account for the unique characteristics of different LLMs and application domains.

\noindent \textbf{Domain-specific RAG design for code-related tasks.} 
Retrieval for code tasks encompasses two primary categories based on retrieved information type: function documentation and similar code snippets. 
Unlike traditional QA tasks where retrieved documents often directly contain required information, code retrieval provides contextual or structural guidance rather than explicit solutions.
This fundamental difference introduces unique challenges in collecting and processing useful knowledge for indexed retrieval databases, as well as in defining and identifying golden documents for retrieval quality evaluation. Despite these challenges' importance, foundational research in these areas remains limited, representing a significant opportunity for future investigation.

\vspace{-10pt}

\subsection{Limitation}
Our exclusive reliance on \textsc{text-embedding-3-small} from OpenAI for retrieval in document number selection (RQ2) and prompting methods (RQ3) studies presents a potential threat to generalizability. 
RAG system behavior may differ when employing alternative retrievers.
However, our use of a mainstream, widely-adopted embedding model suggests that retrieval performance variations would be modest across other advanced retrievers. 
Furthermore, different retrievers primarily introduce variations in the proportions of distracting and golden documents—effects that we systematically examine in RQ1, information completeness and noise are thoroughly studied in RQ2. These design choices ensure our findings remain robust despite this limitation.

\vspace{-12pt}

\section{Related Work}
\label{related_work}

\subsection{Engineering Practices for AI systems}

AI systems are now a fundamental component of modern software infrastructure. For years, the software engineering community has worked to develop better practices that make the engineering, management, deployment, and monitoring of these systems more transparent, reliable, and efficient~\cite{serban2020adoption, CaseStudy}.

Among these efforts, a line of work focuses on studying common bugs and failures in the AI-supporting stack, such as issues in deep learning frameworks~\cite{chen2023toward, tensorflowBugs, islam2019comprehensive}, compilers~\cite{shen2021comprehensive, du2021empirical}, accelerators~\cite{he2023understanding}, mobile platforms~\cite{Deploy2}, and cluster infrastructure~\cite{zhang2020empirical}. By empirically investigating common issues inherent in these commonly used infrastructures, these studies provided suggestions for developers to develop more robust AI systems. Moving to a higher-level software structure, many studies tried to explore better engineering practices at different stages of AI-enabled software, which includes how to address common challenges in DL application deployment~\cite{MLSysChallenges, morovati2023bugs, humbatova2020taxonomy, paleyes2022challenges, Deploy1}, how a team can better collaborate on complex AI system engineering~\cite{nahar2022collaboration}, how to mitigate performance problems in DL systems~\cite{cao2022understanding}, how to better integrate ML components to AI systems~\cite{CloudAPI1, sens2025large} and how to manage ML related assests efficiently~\cite{idowu2021asset}. In contrast to these previous studies, our work focuses on the recently introduced RAG-LLM systems, which may operate with fundamentally different mechanisms and behaviors compared to earlier AI systems.

\subsection{Engineering Practices in RAG System Engineering}
\newcommand{\etal}{\textit{et~al.}}

RAG-based LLMs are complex systems with many design choices that can significantly impact their real-world performance. Several previous studies have examined how different factors and design decisions in RAG systems influence overall outcomes. This includes, but is not limited to, prompt design, retriever selection, reranking algorithm, and database construction. 

Document quality and retrieval factors have received considerable attention in RAG research. Liu~\etal~\cite{lostinthemiddle} examined the influence of document positioning and found that placing golden documents in the middle of the prompt leads to inferior performance. Cuconasu~\etal~\cite{PowerofNoise} analyzed the effects of document type and position on RAG system performance, identifying the negative impact of distracting documents and the potential benefits of noise documents. These studies analyze the effectiveness of factors (i.e. document type, retrieval rank) within RAG systems rather than addressing practical RAG engineering, and are limited to the retrieval phase and focus solely on a single QA task.
The retrieval phase also plays a crucial role in the entire pipeline~\cite{sun2025dynamicrag, searchBestPracticeRAG, wang-etal-2024-rear, park-etal-2025-mirage, yang2025empirical}. As a representative study, Wang~\etal~\cite{searchBestPracticeRAG} compared various design choices in the retrieval phase of RAG systems, such as retrievers and reranking techniques. Parl~\etal~\cite{park-etal-2025-mirage} conducted a comprehensive benchmark of RAG systems, evaluating various retriever-LLM configurations and providing practical guidance based on their findings. Yang~\etal~\cite{yang2025empirical} systematically studied retrieval-augmented frameworks for code generation across three pre-trained models, providing recommendations for different fusion strategies. 
However, these studies predominantly prioritize algorithmic optimizations (e.g., retriever architectures) over the foundational engineering trade-offs determining deployment success. By focusing on isolated components, they overlook antecedent design decisions we investigate: the cost-benefit analysis of whether to deploy RAG, the systemic impact of retrieval volume ($k$), and high-level integration strategies.

In addition to these studies on the design choices of RAG systems, there is also a line of research focused on identifying common pitfalls in RAG system engineering. Barnett~\etal~\cite{SevenFP} identified seven failure points in RAG systems through three case studies from a software engineering perspective. While these studies effectively classify the symptoms of RAG failure (e.g., hallucinations), they do not link them to the upstream engineering decisions that cause them. They lack prescriptive guidance on configuring fundamental parameters to preemptively avoid these pitfalls. Shao~\etal conducted a comprehensive study of 100 open-source applications that incorporate LLMs with RAG support, and identified 18 defect patterns that can potentially degrade the RAG system's functionality and security. In contrast to these works, our study moves beyond component-level benchmarks and defect taxonomies to provide a systematic analysis of the decision-making process itself.

\section{Conclusion}

In this paper, we conduct the first comprehensive empirical study of fundamental RAG deployment decisions, systematically investigating three universal design choices across three LLMs and six datasets spanning question answering and code generation tasks. Our study reveals that effective RAG deployment requires context-aware, strategic decision-making rather than universal adoption. We find that RAG deployment should be highly selective, optimal retrieval volume exhibits strong task-dependent pattern, and prompting methods create orthogonal problem-solving pathways rather than uniform performance enhancements. These results demonstrate that universal RAG strategies are inadequate and that effective systems require careful consideration of task-specific requirements, model characteristics, and knowledge integration methods. Our work offers practical recommendations for key engineering choices and contributes to systematic frameworks for informed RAG development. These contributions support both the software engineering and AI communities in making principled deployment decisions that advance from algorithmic innovations toward successful production implementations.

\section{Acknowledgement}

This paper is supported by National Natural Science Foundation of China (Grant No. 62402183). This work is also supported in part by Canada CIFAR AI Chairs Program, the Natural Sciences and Engineering Research Council of Canada, and the Autoware Foundation.

\bibliographystyle{ACM-Reference-Format}
\bibliography{sample-base}

\appendix

\section*{Appendix}

\subsection*{Detailed Statistical Analysis of Retrieval Scaling}
\label{app:statistical_analysis}

In this appendix, we provide a comprehensive statistical breakdown of the document number experiments reported in Section~\ref{sec:RQ2}. To ensure the robustness of our findings, we applied the following rigorous statistical protocols:

\begin{itemize}
    \item \textbf{False Discovery Rate (FDR) Control:} Significance levels for all McNemar's tests were adjusted using the \textit{Benjamini-Hochberg} procedure to control the False Discovery Rate at $\alpha=0.05$. This correction mitigates the risk of Type I errors arising from multiple pair-wise comparisons.
    \item \textbf{Confidence Intervals:} We report 95\% Wilson Confidence Intervals (CIs) for all accuracy measurements. Unlike standard normal approximation intervals, Wilson intervals provide more accurate coverage for proportions, particularly when performance is near the boundaries (0 or 1).
    \item \textbf{Effect Sizes:} To assess practical significance beyond statistical thresholds, we report both the \textit{Absolute Effect} (arithmetic difference in accuracy) and the \textit{Relative Effect} (percentage improvement relative to the baseline).
\end{itemize}

Table~\ref{tab:appendix_stats} details the pairwise comparisons for consecutive retrieval depths ($k$) across all evaluated models and datasets.

\begin{table*}
\centering
\scriptsize
\caption{Detailed statistical analysis of retrieval scaling. This table reports the absolute accuracy at retrieval depth $k$ and $k'$ with \textbf{95\% Wilson Confidence Intervals} in brackets. We also report the Absolute Effect ($\Delta$), Relative Effect (\%), and the Benjamini-Hochberg FDR-corrected p-values ($p_{adj}$).}
\label{tab:appendix_stats}
\setlength{\tabcolsep}{3.5pt}
\begin{tabular}{llccccc}
\toprule
\textbf{Model / Dataset} & \textbf{Comparison} & \textbf{Acc @ $k$ [95\% CI]} & \textbf{Acc @ $k'$ [95\% CI]} & \textbf{Abs. $\Delta$} & \textbf{Rel. $\Delta$ (\%)} & \textbf{$p_{adj}$ (FDR)} \\
\midrule

\multirow{5}{*}{\textbf{Llama2-13b / NQ}}
& $k=1$ vs $k=3$   & 0.435 [0.413, 0.457] & 0.516 [0.494, 0.538] & +0.081 & +18.6\% & $\mathbf{< 0.001}$ \\
& $k=3$ vs $k=5$   & 0.516 [0.494, 0.538] & 0.544 [0.523, 0.566] & +0.029 & +5.5\%  & $\mathbf{< 0.001}$ \\
& $k=5$ vs $k=10$  & 0.544 [0.523, 0.566] & 0.559 [0.537, 0.581] & +0.015 & +2.7\%  & 0.131 \\
& $k=10$ vs $k=15$ & 0.559 [0.537, 0.581] & 0.552 [0.531, 0.574] & -0.007 & -1.2\%  & 0.535 \\
& $k=15$ vs $k=20$ & 0.552 [0.531, 0.574] & 0.546 [0.525, 0.568] & -0.006 & -1.1\%  & 0.594 \\
\midrule

\multirow{5}{*}{\textbf{Llama2-13b / TriviaQA}}
& $k=1$ vs $k=3$   & 0.732 [0.712, 0.750] & 0.779 [0.761, 0.797] & +0.048 & +6.6\%  & $\mathbf{< 0.001}$ \\
& $k=3$ vs $k=5$   & 0.779 [0.761, 0.797] & 0.801 [0.783, 0.818] & +0.022 & +2.8\%  & $\mathbf{< 0.001}$ \\
& $k=5$ vs $k=10$  & 0.801 [0.783, 0.818] & 0.825 [0.807, 0.841] & +0.023 & +2.9\%  & $0.002^{**}$ \\
& $k=10$ vs $k=15$ & 0.825 [0.807, 0.841] & 0.836 [0.819, 0.852] & +0.012 & +1.4\%  & 0.121 \\
& $k=15$ vs $k=20$ & 0.836 [0.819, 0.852] & 0.835 [0.818, 0.850] & -0.002 & -0.2\%  & 0.926 \\
\midrule

\multirow{5}{*}{\textbf{Llama2-13b / HotpotQA}}
& $k=1$ vs $k=3$   & 0.354 [0.334, 0.376] & 0.441 [0.419, 0.462] & +0.086 & +24.3\% & $\mathbf{< 0.001}$ \\
& $k=3$ vs $k=5$   & 0.441 [0.419, 0.462] & 0.415 [0.394, 0.437] & -0.026 & -5.8\%  & $0.006^{**}$ \\
& $k=5$ vs $k=10$  & 0.415 [0.394, 0.437] & 0.443 [0.421, 0.464] & +0.028 & +6.6\%  & $\mathbf{< 0.001}$ \\
& $k=10$ vs $k=15$ & 0.443 [0.421, 0.464] & 0.436 [0.414, 0.458] & -0.007 & -1.5\%  & 0.535 \\
& $k=15$ vs $k=20$ & 0.436 [0.414, 0.458] & 0.429 [0.407, 0.451] & -0.007 & -1.6\%  & 0.498 \\
\midrule

\multirow{7}{*}{\textbf{GPT-3.5-Turbo / NQ}}
& $k=1$ vs $k=3$   & 0.427 [0.405, 0.449] & 0.500 [0.479, 0.522] & +0.074 & +17.2\% & $\mathbf{< 0.001}$ \\
& $k=3$ vs $k=5$   & 0.500 [0.479, 0.522] & 0.525 [0.503, 0.547] & +0.025 & +4.9\%  & $0.001^{**}$ \\
& $k=5$ vs $k=10$  & 0.525 [0.503, 0.547] & 0.543 [0.521, 0.565] & +0.018 & +3.4\%  & $0.025^{*}$ \\
& $k=10$ vs $k=15$ & 0.543 [0.521, 0.565] & 0.545 [0.524, 0.567] & +0.003 & +0.5\%  & 0.801 \\
& $k=15$ vs $k=20$ & 0.545 [0.524, 0.567] & 0.545 [0.523, 0.567] & -0.001 & -0.1\%  & 1.000 \\
& $k=20$ vs $k=30$ & 0.545 [0.523, 0.567] & 0.549 [0.527, 0.571] & +0.004 & +0.7\%  & 0.610 \\
& $k=30$ vs $k=40$ & 0.549 [0.527, 0.571] & 0.549 [0.527, 0.571] & +0.000 & +0.0\%  & 0.947 \\
\midrule

\multirow{7}{*}{\textbf{GPT-3.5-Turbo / TriviaQA}}
& $k=1$ vs $k=3$   & 0.740 [0.720, 0.759] & 0.776 [0.757, 0.794] & +0.036 & +4.9\%  & $\mathbf{< 0.001}$ \\
& $k=3$ vs $k=5$   & 0.776 [0.757, 0.794] & 0.789 [0.771, 0.806] & +0.013 & +1.7\%  & $0.076^{\dagger}$ \\
& $k=5$ vs $k=10$  & 0.789 [0.771, 0.806] & 0.818 [0.800, 0.834] & +0.029 & +3.6\%  & $\mathbf{< 0.001}$ \\
& $k=10$ vs $k=15$ & 0.818 [0.800, 0.834] & 0.824 [0.806, 0.840] & +0.006 & +0.7\%  & 0.387 \\
& $k=15$ vs $k=20$ & 0.824 [0.806, 0.840] & 0.820 [0.803, 0.836] & -0.004 & -0.4\%  & 0.610 \\
& $k=20$ vs $k=30$ & 0.820 [0.803, 0.836] & 0.828 [0.811, 0.844] & +0.008 & +1.0\%  & 0.255 \\
& $k=30$ vs $k=40$ & 0.828 [0.811, 0.844] & 0.840 [0.823, 0.855] & +0.012 & +1.4\%  & $0.034^{*}$ \\
\midrule

\multirow{7}{*}{\textbf{GPT-3.5-Turbo / HotpotQA}}
& $k=1$ vs $k=3$   & 0.346 [0.325, 0.367] & 0.391 [0.369, 0.412] & +0.045 & +12.9\% & $\mathbf{< 0.001}$ \\
& $k=3$ vs $k=5$   & 0.391 [0.369, 0.412] & 0.407 [0.386, 0.429] & +0.017 & +4.2\%  & $0.055^{\dagger}$ \\
& $k=5$ vs $k=10$  & 0.407 [0.386, 0.429] & 0.423 [0.402, 0.445] & +0.016 & +3.9\%  & $0.071^{\dagger}$ \\
& $k=10$ vs $k=15$ & 0.423 [0.402, 0.445] & 0.428 [0.406, 0.450] & +0.005 & +1.2\%  & 0.589 \\
& $k=15$ vs $k=20$ & 0.428 [0.406, 0.450] & 0.427 [0.405, 0.449] & -0.001 & -0.2\%  & 0.947 \\
& $k=20$ vs $k=30$ & 0.427 [0.405, 0.449] & 0.436 [0.414, 0.458] & +0.009 & +2.1\%  & 0.255 \\
& $k=30$ vs $k=40$ & 0.436 [0.414, 0.458] & 0.438 [0.416, 0.459] & +0.002 & +0.3\%  & 0.918 \\
\midrule

\multirow{7}{*}{\textbf{GPT-4o-mini / NQ}}
& $k=1$ vs $k=3$   & 0.495 [0.473, 0.517] & 0.550 [0.528, 0.572] & +0.055 & +11.1\% & $\mathbf{< 0.001}$ \\
& $k=3$ vs $k=5$   & 0.550 [0.528, 0.572] & 0.587 [0.565, 0.608] & +0.037 & +6.7\%  & $\mathbf{< 0.001}$ \\
& $k=5$ vs $k=10$  & 0.587 [0.565, 0.608] & 0.602 [0.580, 0.623] & +0.015 & +2.6\%  & $0.039^{*}$ \\
& $k=10$ vs $k=15$ & 0.602 [0.580, 0.623] & 0.613 [0.591, 0.634] & +0.011 & +1.8\%  & $0.088^{\dagger}$ \\
& $k=15$ vs $k=20$ & 0.613 [0.591, 0.634] & 0.618 [0.596, 0.639] & +0.005 & +0.7\%  & 0.535 \\
& $k=20$ vs $k=30$ & 0.618 [0.596, 0.639] & 0.624 [0.602, 0.644] & +0.006 & +1.0\%  & 0.407 \\
& $k=30$ vs $k=40$ & 0.624 [0.602, 0.644] & 0.627 [0.606, 0.648] & +0.004 & +0.6\%  & 0.498 \\
\midrule

\multirow{7}{*}{\textbf{GPT-4o-mini / TriviaQA}}
& $k=1$ vs $k=3$   & 0.779 [0.761, 0.797] & 0.815 [0.797, 0.831] & +0.036 & +4.6\%  & $\mathbf{< 0.001}$ \\
& $k=3$ vs $k=5$   & 0.815 [0.797, 0.831] & 0.823 [0.805, 0.839] & +0.008 & +0.9\%  & 0.297 \\
& $k=5$ vs $k=10$  & 0.823 [0.805, 0.839] & 0.841 [0.824, 0.856] & +0.019 & +2.2\%  & $0.002^{**}$ \\
& $k=10$ vs $k=15$ & 0.841 [0.824, 0.856] & 0.850 [0.833, 0.864] & +0.009 & +1.0\%  & 0.112 \\
& $k=15$ vs $k=20$ & 0.850 [0.833, 0.864] & 0.857 [0.841, 0.872] & +0.008 & +0.9\%  & 0.112 \\
& $k=20$ vs $k=30$ & 0.857 [0.841, 0.872] & 0.861 [0.845, 0.875] & +0.004 & +0.4\%  & 0.590 \\
& $k=30$ vs $k=40$ & 0.861 [0.845, 0.875] & 0.865 [0.849, 0.879] & +0.004 & +0.5\%  & 0.498 \\
\midrule

\multirow{7}{*}{\textbf{GPT-4o-mini / HotpotQA}}
& $k=1$ vs $k=3$   & 0.413 [0.392, 0.435] & 0.440 [0.418, 0.461] & +0.027 & +6.4\%  & $0.008^{**}$ \\
& $k=3$ vs $k=5$   & 0.440 [0.418, 0.461] & 0.464 [0.442, 0.486] & +0.025 & +5.6\%  & $\mathbf{< 0.001}$ \\
& $k=5$ vs $k=10$  & 0.464 [0.442, 0.486] & 0.494 [0.472, 0.516] & +0.030 & +6.5\%  & $\mathbf{< 0.001}$ \\
& $k=10$ vs $k=15$ & 0.494 [0.472, 0.516] & 0.497 [0.476, 0.519] & +0.004 & +0.7\%  & 0.684 \\
& $k=15$ vs $k=20$ & 0.497 [0.476, 0.519] & 0.500 [0.479, 0.522] & +0.003 & +0.6\%  & 0.701 \\
& $k=20$ vs $k=30$ & 0.500 [0.479, 0.522] & 0.515 [0.494, 0.537] & +0.015 & +3.0\%  & $0.034^{*}$ \\
& $k=30$ vs $k=40$ & 0.515 [0.494, 0.537] & 0.524 [0.503, 0.546] & +0.009 & +1.7\%  & 0.192 \\

\bottomrule
\end{tabular}
\vspace{5pt}
\end{table*}

\subsection*{Statistical Significance of Perplexity Variance}
\label{app:ppl_analysis}

In this appendix, we analyze the statistical significance of Perplexity (PPL) variance to determine if changes in retrieval depth ($k$) induce meaningful shifts in the model's predictive distribution. Table~\ref{tab:ppl_significance} details the pairwise comparisons across datasets and models as the document number $k$ increases. While the absolute magnitude of PPL variations often appears subtle, this analysis demonstrates that these shifts remain statistically significant, particularly in the majority of QA dataset cases.

\begin{table*}[t]
\centering
\scriptsize
\caption{Perplexity (PPL) variance significance with increasing document number across models and datasets. Legend: \H\H: Highly Significant ($p < 0.01$), \H: Significant ($p < 0.05$), \M: Marginal ($p < 0.1$), \N: Not Significant, -: Not Evaluated.}
\label{tab:ppl_significance}
\setlength{\tabcolsep}{1.8pt} 
\begin{tabular}{llcccccccccccc}
\toprule
& & \multicolumn{12}{c}{\textbf{Retrieval Context Scaling (Step Comparison)}} \\
\cmidrule(lr){3-14}
\textbf{Model Family} & \textbf{Dataset} & \textbf{1-3} & \textbf{3-5} & \textbf{5-7} & \textbf{5-10} & \textbf{7-10} & \textbf{10-13} & \textbf{10-15} & \textbf{13-16} & \textbf{15-20} & \textbf{16-20} & \textbf{20-30} & \textbf{30-40} \\
\midrule

\multirow{6}{*}{\shortstack[l]{\textbf{Llama2-13B} / \\ \textbf{CodeLlama}}} 
 & NQ & \H\H & \H\H & - & \H\H & - & - & \N & - & \H\H & - & - & - \\
 & TriviaQA & \H\H & \H\H & - & \H\H & - & - & \H\H & - & \H\H & - & - & - \\
 & HotpotQA & \H\H & \H\H & - & \N & - & - & \H\H & - & \H\H & - & - & - \\
 & CoNaLa & \N & \N & \N & - & \N & \N & - & \N & - & \N & - & - \\
 & DS1000 & \H\H & \N & \N & - & \N & \H & - & \N & - & \N & - & - \\
 & PNE & \H\H & \N & \M & - & \N & \N & - & \N & - & \N & - & - \\
\midrule

\multirow{6}{*}{\textbf{GPT-3.5-Turbo}} 
 & NQ & \N & \H & - & \H\H & - & - & \H & - & \H & - & \H\H & \H\H \\
 & TriviaQA & \H\H & \M & - & \N & - & - & \H\H & - & \H & - & \H\H & \H\H \\
 & HotpotQA & \M & \H & - & \N & - & - & \H & - & \N & - & \H & \H \\
 & CoNaLa & \M & \N & \N & - & \N & \N & - & \N & - & \N & - & - \\
 & DS1000 & \H & \N & \N & - & \N & \N & - & \N & - & \N & - & - \\
 & PNE & \M & \N & \N & - & \N & \N & - & \N & - & \N & - & - \\
\midrule

\multirow{6}{*}{\textbf{GPT-4o-mini}} 
 & NQ & \N & \H & - & \H\H & - & - & \N & - & \H\H & - & \N & \M \\
 & TriviaQA & \N & \M & - & \H & - & - & \H & - & \N & - & \H & \N \\
 & HotpotQA & \H\H & \N & - & \H & - & - & \N & - & \M & - & \N & \N \\
 & CoNaLa & \N & \N & \N & - & \N & \N & - & \N & - & \N & - & - \\
 & DS1000 & \M & \N & \N & - & \N & \N & - & \N & - & \N & - & - \\
 & PNE & \H\H & \N & \N & - & \H\H & \H\H & - & \H & - & \N & - & - \\

\bottomrule
\end{tabular}
\end{table*}

\begin{figure*}[ht]
    \centering
    \includegraphics[width=0.9\textwidth]{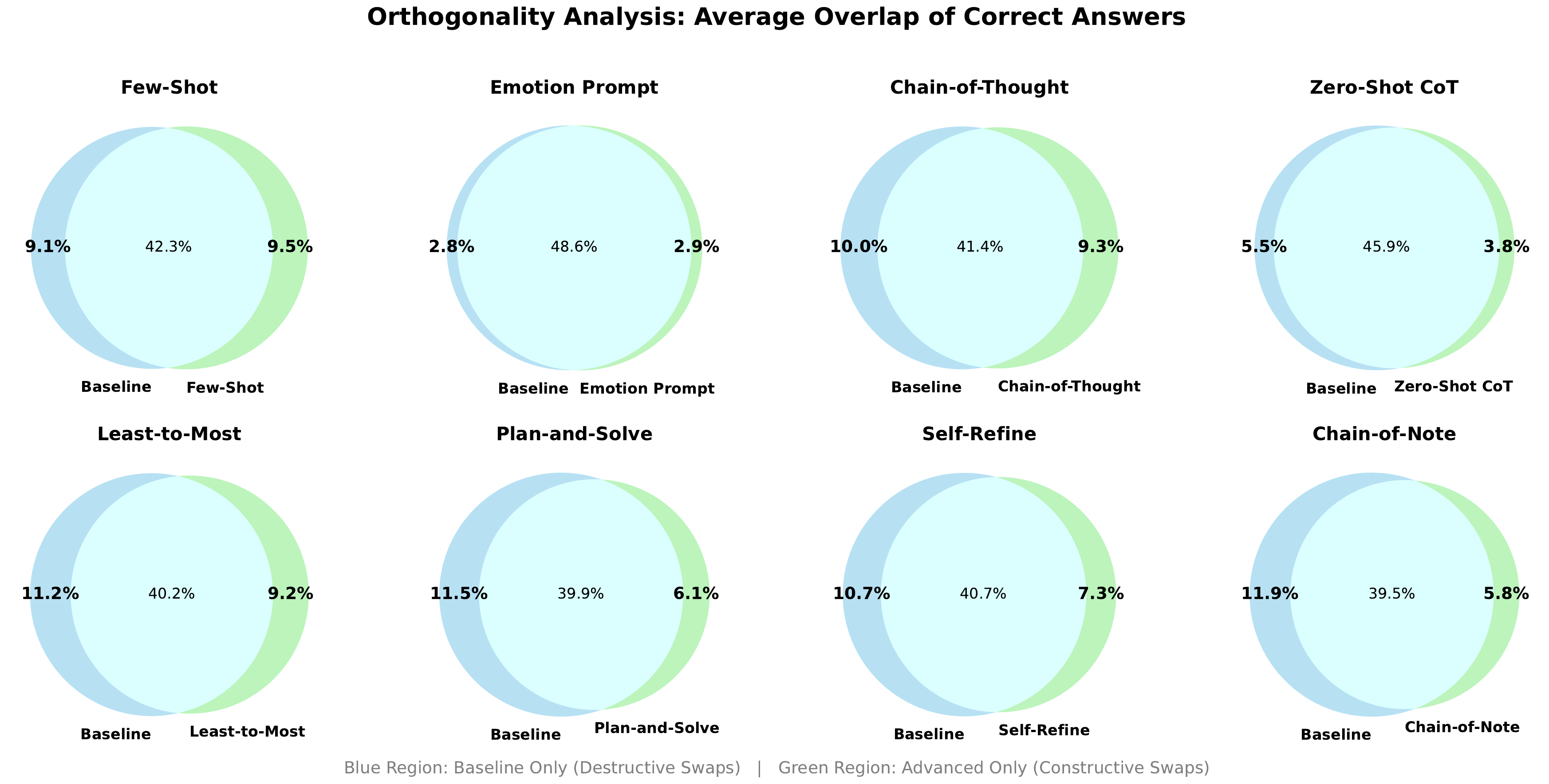}
    \caption{Visualization of orthogonal effects and swap rates. Venn diagrams illustrating solution set overlaps between the Baseline (Left/Blue) and Advanced methods (Right/Green), averaged across all 3 models and 6 datasets. Non-overlapping regions represent uniquely solved queries, demonstrating that advanced strategies cover orthogonal problem spaces rather than strictly superseding the baseline.}
    \label{fig:venn_orthogonal}
\end{figure*}

\subsection*{Visualization of Orthogonal Effects}
\label{app:orthogonality}

To substantiate the ``orthogonal solves'' claim presented in Section~\ref{sec:RQ3}, we further visualize the specific solution overlaps between prompting methods, shown in Figure~\ref{fig:venn_orthogonal}. This figure displays the average intersection between the baseline prompt and each advanced prompting method across all models and datasets.

The visualization confirms that performance differences are not merely additive. For methods like \textit{Few-Shot} and \textit{Chain-of-Thought}, we observe substantial non-overlapping regions: the blue region (Baseline-only correct) and the green region (advanced-only correct) are both significant (e.g., $9\text{--}10\%$ for CoT and Few-Shot). 

\subsection*{Bootstrap Stability Analysis of Orthogonality}
\label{app:bootstrap}

To verify the statistical stability of the ``orthogonal solves'' phenomenon in Section~\ref{sec:RQ3}, we performed a bootstrap analysis with $N=1,000$ iterations. Table~\ref{tab:bootstrap_complete} presents the mean percentages and 95\% Confidence Intervals (CIs) for samples that are uniquely correct in the Baseline method versus uniquely correct in the Advanced method.

While we observe varying degrees of stability—ranging from minimal CIs in lower-impact methods (e.g., Llama2-13B with \textit{Emotion Prompting}) to wider intervals in high-variance strategies (e.g., Llama2-13B with \textit{Least-to-Most})—the overall trend validates the orthogonality hypothesis. In many key scenarios, the lower bound of the unique solution rate exceeds 10\%. Furthermore, in the majority of cases, the lower bounds for both ``Baseline-Only'' and ``Advanced-Only'' sets surpass 5\%. This statistically confirms that the orthogonal effect is a robust property: advanced methods consistently solve a distinct, non-overlapping subset of problems rather than merely refining the baseline's existing answers.

\begin{table*}[h]
\centering
\scriptsize
\renewcommand{\arraystretch}{1.25} 
\caption{Bootstrap analysis of prediction distribution.
To substantiate the ``orthogonal solves'' claim, this table details the stability of solution overlaps. 
Values represent the percentage of samples correct only in the baseline (Top) versus samples correct only in the advanced method (Bottom). 
Format: Mean [95\% Confidence Interval], calculated via bootstrap resampling ($N=1,000$). 
The narrow confidence intervals confirm that the specific subsets of queries uniquely solved by each method are statistically stable phenomena rather than random variance.}
\label{tab:bootstrap_complete}
\setlength{\tabcolsep}{2.5pt} 
\begin{tabular}{l|ccc|ccc}
\toprule
\multirow{2}{*}{\textbf{Method}} & \multicolumn{3}{c}{\textbf{QA Tasks}} & \multicolumn{3}{c}{\textbf{Code Tasks}} \\
\cmidrule(lr){2-4} \cmidrule(lr){5-7}
 & \textbf{NQ} & \textbf{TriviaQA} & \textbf{HotpotQA} & \textbf{CoNaLa} & \textbf{DS1000} & \textbf{PNE} \\
\midrule

\multicolumn{7}{c}{\cellcolor{gray!10}\textbf{Llama2-13B / CodeLlama-13B}} \\
\midrule
Few-Shot & \makecell{15.1 [13.6--16.8] \\ 4.0 [3.2--4.9]} & \makecell{8.2 [7.0--9.4] \\ 2.1 [1.6--2.8]} & \makecell{12.9 [11.5--14.5] \\ 3.8 [2.9--4.7]} & \makecell{8.4 [2.4--14.3] \\ 22.7 [14.3--32.1]} & \makecell{3.9 [1.3--7.0] \\ 13.3 [8.3--18.5]} & \makecell{7.7 [4.2--12.0] \\ 20.9 [15.0--27.5]} \\
\addlinespace[3pt]
Emotion & \makecell{2.0 [1.4--2.7] \\ 1.8 [1.3--2.4]} & \makecell{1.5 [1.0--2.1] \\ 0.5 [0.3--0.9]} & \makecell{3.3 [2.5--4.1] \\ 1.3 [0.8--1.8]} & \makecell{2.4 [0.0--6.0] \\ 0.0 [0.0--0.0]} & \makecell{5.1 [1.9--8.9] \\ 2.5 [0.6--5.1]} & \makecell{2.5 [0.6--4.8] \\ 4.1 [1.8--7.2]} \\
\addlinespace[3pt]
CoT & \makecell{14.4 [12.9--15.9] \\ 4.0 [3.2--4.9]} & \makecell{8.1 [6.9--9.2] \\ 1.6 [1.0--2.2]} & \makecell{13.4 [11.9--14.9] \\ 6.0 [5.1--7.1]} & \makecell{5.9 [1.2--11.9] \\ 26.0 [16.7--35.7]} & \makecell{6.4 [2.5--10.8] \\ 11.5 [7.0--17.2]} & \makecell{15.5 [10.2--21.6] \\ 16.2 [10.8--22.2]} \\
\addlinespace[3pt]
Zero-Shot CoT & \makecell{5.8 [4.8--6.9] \\ 4.1 [3.3--5.0]} & \makecell{3.3 [2.6--4.2] \\ 2.0 [1.4--2.6]} & \makecell{6.4 [5.4--7.4] \\ 4.5 [3.6--5.5]} & \makecell{1.2 [0.0--4.8] \\ 3.7 [0.0--8.3]} & \makecell{9.6 [5.7--14.6] \\ 3.9 [1.3--7.0]} & \makecell{19.8 [13.8--25.7] \\ 7.1 [3.6--11.4]} \\
\addlinespace[3pt]
Least-to-Most & \makecell{13.0 [11.5--14.5] \\ 5.4 [4.5--6.4]} & \makecell{7.1 [5.9--8.2] \\ 2.6 [1.9--3.3]} & \makecell{11.2 [9.8--12.6] \\ 7.0 [5.9--8.3]} & \makecell{11.9 [6.0--19.0] \\ 30.9 [21.4--40.5]} & \makecell{7.1 [3.8--11.5] \\ 9.0 [5.1--14.0]} & \makecell{21.5 [15.6--28.1] \\ 13.2 [8.4--18.6]} \\
\addlinespace[3pt]
Plan-and-Solve & \makecell{21.6 [19.8--23.3] \\ 5.6 [4.8--6.6]} & \makecell{28.4 [26.5--30.3] \\ 3.5 [2.8--4.4]} & \makecell{23.4 [21.7--25.3] \\ 5.1 [4.1--6.1]} & \makecell{11.9 [6.0--19.0] \\ 11.9 [4.8--19.0]} & \makecell{10.9 [6.4--15.9] \\ 1.9 [0.0--4.5]} & \makecell{25.7 [19.2--32.3] \\ 5.9 [2.4--9.6]} \\
\addlinespace[3pt]
Self-Refine & \makecell{17.2 [15.6--19.0] \\ 4.5 [3.6--5.4]} & \makecell{11.9 [10.5--13.3] \\ 2.8 [2.2--3.6]} & \makecell{19.5 [17.8--21.2] \\ 5.0 [4.1--5.9]} & \makecell{3.6 [0.0--8.3] \\ 15.4 [8.3--23.8]} & \makecell{5.7 [2.5--9.6] \\ 1.9 [0.0--4.5]} & \makecell{14.4 [9.6--19.8] \\ 6.6 [3.0--10.8]} \\
\addlinespace[3pt]
Chain-of-Note & \makecell{12.5 [11.1--14.0] \\ 4.9 [4.0--5.8]} & \makecell{9.6 [8.3--10.9] \\ 3.1 [2.4--3.9]} & \makecell{19.1 [17.4--20.7] \\ 5.2 [4.3--6.2]} & \makecell{14.1 [7.1--22.6] \\ 4.8 [1.2--9.5]} & \makecell{8.2 [4.5--12.7] \\ 1.3 [0.0--3.2]} & \makecell{24.6 [18.6--31.1] \\ 4.2 [1.2--7.2]} \\

\midrule
\multicolumn{7}{c}{\cellcolor{gray!10}\textbf{GPT-3.5-Turbo}} \\
\midrule
Few-Shot & \makecell{10.5 [9.1--11.9] \\ 5.4 [4.5--6.4]} & \makecell{7.7 [6.6--8.8] \\ 3.5 [2.7--4.4]} & \makecell{7.1 [6.0--8.2] \\ 5.5 [4.5--6.5]} & \makecell{10.7 [4.8--16.7] \\ 24.9 [16.7--34.5]} & \makecell{12.8 [7.6--18.5] \\ 10.9 [6.4--15.3]} & \makecell{4.2 [1.8--7.2] \\ 13.9 [9.0--19.8]} \\
\addlinespace[3pt]
Emotion & \makecell{2.6 [2.0--3.4] \\ 2.5 [1.8--3.2]} & \makecell{1.4 [1.0--2.0] \\ 1.3 [0.8--1.8]} & \makecell{3.0 [2.2--3.8] \\ 2.6 [1.9--3.4]} & \makecell{4.9 [1.2--9.5] \\ 10.7 [4.8--17.9]} & \makecell{3.9 [1.3--7.0] \\ 7.7 [3.8--12.1]} & \makecell{4.2 [1.8--7.2] \\ 2.4 [0.6--4.8]} \\
\addlinespace[3pt]
CoT & \makecell{8.6 [7.4--9.9] \\ 5.3 [4.3--6.4]} & \makecell{6.1 [5.0--7.1] \\ 4.1 [3.3--4.9]} & \makecell{7.6 [6.6--8.8] \\ 8.8 [7.5--10.1]} & \makecell{10.6 [4.8--17.9] \\ 21.4 [13.1--31.0]} & \makecell{12.1 [7.6--17.2] \\ 7.6 [3.8--12.1]} & \makecell{8.3 [4.2--12.6] \\ 12.5 [7.8--17.4]} \\
\addlinespace[3pt]
Zero-Shot CoT & \makecell{2.7 [2.0--3.4] \\ 1.5 [1.0--2.0]} & \makecell{2.0 [1.5--2.7] \\ 1.5 [1.0--2.0]} & \makecell{3.4 [2.6--4.2] \\ 2.7 [2.0--3.4]} & \makecell{4.7 [1.2--9.5] \\ 5.9 [1.2--11.9]} & \makecell{4.4 [1.3--7.6] \\ 6.4 [2.5--10.2]} & \makecell{7.2 [3.6--11.4] \\ 4.2 [1.8--7.8]} \\
\addlinespace[3pt]
Least-to-Most & \makecell{8.4 [7.4--9.7] \\ 7.0 [5.8--8.1]} & \makecell{4.9 [4.1--5.9] \\ 4.9 [4.0--5.9]} & \makecell{8.2 [7.0--9.5] \\ 10.1 [8.8--11.4]} & \makecell{10.8 [4.8--17.9] \\ 15.5 [8.3--23.8]} & \makecell{12.8 [8.3--17.8] \\ 8.4 [4.5--12.7]} & \makecell{12.0 [6.6--17.4] \\ 11.9 [7.2--17.4]} \\
\addlinespace[3pt]
Plan-and-Solve & \makecell{11.3 [10.0--12.6] \\ 6.5 [5.4--7.6]} & \makecell{11.0 [9.7--12.4] \\ 4.2 [3.3--5.1]} & \makecell{8.4 [7.2--9.7] \\ 12.2 [10.8--13.7]} & \makecell{3.6 [0.0--8.3] \\ 10.8 [4.8--17.9]} & \makecell{7.0 [3.8--10.8] \\ 6.4 [3.2--10.2]} & \makecell{6.0 [3.0--10.2] \\ 4.8 [1.8--8.4]} \\
\addlinespace[3pt]
Self-Refine & \makecell{8.1 [6.9--9.4] \\ 6.9 [5.8--8.0]} & \makecell{5.9 [5.0--7.0] \\ 5.0 [4.1--6.0]} & \makecell{8.6 [7.4--9.9] \\ 6.9 [5.8--8.0]} & \makecell{10.7 [4.8--17.9] \\ 20.2 [13.1--28.6]} & \makecell{9.0 [5.1--13.4] \\ 7.6 [3.8--12.1]} & \makecell{13.2 [8.4--18.6] \\ 13.9 [9.0--19.2]} \\
\addlinespace[3pt]
Chain-of-Note & \makecell{6.7 [5.6--7.9] \\ 7.8 [6.6--9.0]} & \makecell{7.1 [6.1--8.3] \\ 4.2 [3.4--5.1]} & \makecell{8.7 [7.4--10.1] \\ 8.1 [6.9--9.2]} & \makecell{7.1 [2.4--13.1] \\ 14.3 [7.1--21.5]} & \makecell{13.4 [8.3--19.1] \\ 6.4 [3.2--10.2]} & \makecell{16.8 [11.4--22.8] \\ 9.1 [4.8--13.8]} \\

\midrule
\multicolumn{7}{c}{\cellcolor{gray!10}\textbf{GPT-4o-mini}} \\
\midrule
Few-Shot & \makecell{5.6 [4.7--6.7] \\ 3.4 [2.7--4.2]} & \makecell{1.6 [1.0--2.1] \\ 3.4 [2.6--4.2]} & \makecell{8.0 [6.7--9.2] \\ 3.3 [2.6--4.1]} & \makecell{7.3 [2.4--13.1] \\ 15.3 [8.3--23.8]} & \makecell{29.8 [22.3--37.6] \\ 8.9 [4.5--13.4]} & \makecell{3.0 [0.6--6.0] \\ 6.7 [3.0--10.8]} \\
\addlinespace[3pt]
Emotion & \makecell{1.2 [0.8--1.8] \\ 2.1 [1.5--2.8]} & \makecell{0.8 [0.4--1.3] \\ 0.9 [0.5--1.4]} & \makecell{1.2 [0.8--1.7] \\ 2.4 [1.8--3.1]} & \makecell{0.0 [0.0--0.0] \\ 4.6 [1.2--9.5]} & \makecell{9.6 [5.1--14.0] \\ 3.1 [0.6--6.4]} & \makecell{1.8 [0.0--4.2] \\ 1.8 [0.0--4.2]} \\
\addlinespace[3pt]
CoT & \makecell{7.3 [6.2--8.5] \\ 3.5 [2.7--4.3]} & \makecell{2.8 [2.1--3.6] \\ 2.4 [1.7--3.1]} & \makecell{9.0 [7.8--10.3] \\ 5.8 [4.9--6.9]} & \makecell{7.3 [2.4--13.1] \\ 12.0 [6.0--19.0]} & \makecell{33.7 [26.8--40.8] \\ 8.3 [4.4--12.7]} & \makecell{3.0 [0.6--6.0] \\ 10.0 [6.0--15.0]} \\
\addlinespace[3pt]
Zero-Shot CoT & \makecell{1.4 [0.9--2.0] \\ 2.3 [1.7--2.9]} & \makecell{1.1 [0.7--1.6] \\ 1.1 [0.7--1.6]} & \makecell{6.5 [5.4--7.6] \\ 5.8 [4.8--6.8]} & \makecell{1.2 [0.0--3.6] \\ 4.7 [1.2--9.5]} & \makecell{11.5 [6.4--15.9] \\ 4.4 [1.9--8.3]} & \makecell{7.1 [3.6--10.8] \\ 3.6 [1.2--6.6]} \\
\addlinespace[3pt]
Least-to-Most & \makecell{9.0 [7.9--10.3] \\ 4.1 [3.3--5.1]} & \makecell{3.0 [2.2--3.7] \\ 3.3 [2.5--4.1]} & \makecell{8.5 [7.3--9.9] \\ 7.2 [6.1--8.5]} & \makecell{7.2 [2.4--13.1] \\ 6.0 [1.2--10.7]} & \makecell{31.6 [24.8--38.9] \\ 8.2 [3.8--12.7]} & \makecell{3.5 [1.2--6.6] \\ 10.3 [5.4--15.0]} \\
\addlinespace[3pt]
Plan-and-Solve & \makecell{7.8 [6.6--9.0] \\ 4.8 [3.9--5.7]} & \makecell{3.9 [3.0--4.7] \\ 4.0 [3.1--5.0]} & \makecell{7.7 [6.6--8.8] \\ 8.7 [7.4--9.9]} & \makecell{2.3 [0.0--6.0] \\ 6.0 [2.4--11.9]} & \makecell{13.4 [8.9--18.5] \\ 7.0 [3.2--11.5]} & \makecell{7.8 [4.2--12.0] \\ 4.1 [1.2--7.2]} \\
\addlinespace[3pt]
Self-Refine & \makecell{2.5 [1.8--3.3] \\ 4.5 [3.6--5.5]} & \makecell{1.6 [1.0--2.2] \\ 3.0 [2.2--3.8]} & \makecell{4.3 [3.4--5.3] \\ 4.1 [3.2--4.9]} & \makecell{14.5 [7.1--22.6] \\ 8.2 [2.4--14.3]} & \makecell{33.9 [26.7--40.8] \\ 5.8 [2.5--9.6]} & \makecell{8.9 [4.8--13.2] \\ 9.6 [5.4--14.4]} \\
\addlinespace[3pt]
Chain-of-Note & \makecell{8.0 [6.7--9.2] \\ 3.9 [3.1--4.8]} & \makecell{2.4 [1.8--3.1] \\ 3.6 [2.8--4.5]} & \makecell{6.9 [5.8--8.0] \\ 7.0 [5.9--8.1]} & \makecell{14.2 [7.1--22.6] \\ 3.6 [0.0--8.3]} & \makecell{25.0 [18.5--31.8] \\ 7.0 [3.2--10.8]} & \makecell{9.7 [6.0--14.4] \\ 5.4 [2.4--9.0]} \\

\bottomrule
\end{tabular}
\end{table*}

\end{document}